\documentstyle[10pt]{article}
\newcommand{\gea}{\mbox{$\Gamma$}}
\newcommand{\ef}{\mbox{$\cal{W}$}}
\newcommand{\mfi}[1]{\mbox{ $\tilde{\varphi}^{#1}$ }}
\newcommand{\rx}{\mbox{ $\bar{x}$}}
\newcommand{\rp}{\mbox{ $\bar{p}$}}

\makeatletter                    
\@addtoreset{equation}{section}  
\makeatother                     

\begin{document}
\begin{titlepage}
\begin{flushright}
22-12-1994 \ / \ HD-TVP-94-22
\end{flushright}
\vskip 3cm
\begin{center}
{\Large\bf $\lambda \varphi ^{4}$ non-equilibrium
dynamics and kinetic field theory \footnote{Supported in part by
the Deutsche Forschungs Gemeinschaft, grant number Hu 233/4-3.}}
\vskip 1cm
\mbox{F.M.C.\ Witte
and S.P.\ Klevansky} \\
{
\it
Institut f\"ur Theoretische Physik,\
Universit\"at Heidelberg\\
Philosophenweg  19, \ 69120  Heidelberg \\
Germany}
\end{center}
\date{DECEMBER 1994}
\vskip 1.5cm
\begin{abstract}
\footnotesize
Off-shellness and inhomogeneities in the non-equilibrium
dynamics of the $\lambda \varphi^{4}$ model are studied.
The non equilibrium field theory is developed in the
Closed Time Path formalism, which is in turn reformulated
as a kinetic field theory, in terms of a set of kinetic
equations for the 2-point Green function, derived from
a generalized effective action functional.
We take into account all initial correlations up to the
4-point functions.
It is shown that the model displays an SO(1,1) symmetry
broken by interactions and initial conditions.
The divergence of the corresponding Noether current is
identified as a particle creation/annihilation density.
The broken Ward-Takahashi relations for the SO(1,1)
symmetry are derived.
They constitute a set of generalized kinetic equations
for the general n-point functions.
Energy-momentum conservation is demonstrated to hold
from the fully interacting transport equations,
and we discuss the effect of a gradient expansion and
its non-locality.
As an application, we study an explicit realization
of inhomogeneities and non-equilibrium conditions
in the free field model, for which we give the
general solution and analyze it in a particular case.
The identification of the Casimir effect in our
solution illustrates the intimate connection between
inhomogeneities and off-shellness.
\end{abstract}

\end{titlepage}

\section{Introduction}

\par
\indent
Phase transitions play an important role
in the study of physical systems under extreme conditions.
In particle physics, various kinds of phase
transitions are of interest and the use of
equilibrium thermodynamics may be inadequate.
For example,
this is so in the current descriptions
of the initial stages of heavy ion collisions,
in which,
heuristically,
the onset of deconfinement
is believed to give rise to the formation of a
quark gluon plasma \cite{Kap}.
Two further examples that one may speculate upon,
are the electroweak phase transition in the early universe
\cite{ewk}, and the formation of
disaligned chiral condensates (dcc) in high energy
hadronic collisions \cite{Bjor},
such as $\pi \ p $ scattering.

\par
The closed time\-path (CTP) for\-ma\-lism,
ori\-gina\-ting from the work of Schwinger \cite{Schwing},
Keldysh \cite {Kel}, and Vernon and Feynman \cite{Vfeyn},
is suitable for analyzing non-equilibrium systems.
It takes one beyond the more conventional approaches
used to describe small perturbations
about thermal equilibrium, such as
thermo field dynamics \cite{tdf},
the Niemi-Semenoff methods \cite{NS} and
the real time finite temperature formalism
of Dolan and Jackiw \cite{DoJa}.
The CTP formalism is not restricted
to quasi homogeneous systems close
to equilibrium,
although in practice,
to date,
it has been difficult to implement
non-equilibrium theory in inhomogeneous systems.
The full power of the CTP formalism
is unfolded within the functional integral
representation \cite{Chin,Hu}.
Recently  \cite{Cal,Klei} an
effective action principle,
that leads to a hierarchy of
Schwinger-Dyson equations for the Green functions,
has been formulated.
These equations can be reformulated
in terms of a transport theory,
or kinetic field theory,
for the 2-point functions.
Such a field theoretic formalism can be shown
to have a Boltzmann equation as its
quasi-classical limit \cite{Cal}.
While the CTP non equilibrium formalism
is theoretically aesthetic and complete,
it is difficult to implement practically,
and has only recently become a subject of
detailed study.
Transport theories in general constitute a topic
whose quantum field theoretic  content is difficult
to understand on the basis of quasi-classical models.
We briefly summarize some applications
of transport theory that have been developed and
are currently of interest.
We do so without attempting to be complete.
\par
Non-relativistic kinetic equations
have become a common tool in analyzing
heavy ion collisions numerically \cite{Dan}.
Semi-classical,
but relativistic,
approaches based on quantum
hadron dynamics have been put forward
for the same purpose \cite{Blat}.
A lot of work has been done
on the formulation of a quark-gluon transport
theory based on the
operator formalism
of quantum field theory \cite{Hei}.
In order to study the restoration
of chiral symmetry in relativistic heavy
ion collisions,
a transport theory for the Nambu-Jona-Lasinio model,
that contains quarks as its fundamental degrees of freedom,
has been derived in the framework of the CTP formalism \cite{Wil}.
Due to the difficulties inherent in this particular problem
the equations of motion that are obtained have however
only been solved in the mean field approximation,
in order to study the influence
of the phase transition on the expansion
of a QGP \cite{Aich}, and the fluctuations
in a quark-antiquark plasma \cite{Woi}.
The CTP formalism has also been recently applied
\cite{Mot} to the O(4) linear sigma model
to investigate dcc formation.
In all of the studies that have been mentioned,
the systems have been  required to be quasi
uniform and/or the particles to be on mass-shell.
These restrictions have simply been necessitated
by the dificulty of the problem at hand,
since they give rise to considerable simplifications
in the formalism when applied.
On the other hand, the lifting of these requirements
may in some cases be important.
The mass-shell restriction may suppress
important quantum effects.
This is illustrated in the calculation
of Eisenberg \cite{Eis}
showing that the spontaneous creation
of particles from a constant background
field cannot be described in this way.

\par
In order to obtain a better understanding
of the importance of off-shell
behaviour and spatial inhomogeneities,
it is useful to examine a simple model.
To this end, we shall focus on the well-\-known
scalar $\lambda \varphi ^{4}$ model in
this paper and study the CTP formalism and
associated transport theory for this model.
In doing so we shall endeavour to remove the
on-shell restriction, and also investigate
a scenario which is far from equilibrium.
Of technical importance is the proof that
energy and momentum are conserved
in the presence of interactions,
which we also demonstrate.
\par
Since our analysis is performed
in the setting of kinetic theory,
we discuss this briefly here.
If interactions are short ranged,
as is the case for the $\lambda \varphi ^{4}$
point-like interaction,
one expects naively that the quantum field may,
above some length-scale $\ell_{f}$,
be conceived as a distribution of free particles,
each of which temporarily goes off-shell
while undergoing an interaction, during a time $\tau_{c}$ .
For this picture to have some validity, one has
\begin{equation}
  \tau_{c} \ll \ell_{f} \ .
\end{equation}
In that case, the time dependent
particle density $n(\vec{x},t)$ satisfies
\begin{equation}
  n(\vec{x} , t) \ll  \ell_{f}^{-3}
\end{equation}
in describing the system.
Due to particle creation and annihilation
$n(\vec{x},t)$ will,
in general,
have a time dependent normalization.
\par
This separation of the system into
macroscopic length scales above $\ell_{f}$ ,
and microscopic length scales below $\ell_{f}$,
is at the same time a statement about the effective
interaction,
and hence the effective coupling
of the particles at different energy scales.
It is therefore important to consider the
behavior of the running coupling as we move
from one length scale to another.
For example,
in QCD at high energies,
the quark and gluon
degrees of freedom are relevant,
whereas at low energies these remain confined and
their hadronic bound states can be treated as fundamental.
In this case,
the separation in length scales
is accompanied by a separation of
physical degrees of freedom.
Such a phenomenon is difficult to implement
in a two-scale transport theory.
A two-scale approach is also not consistent with
the running of the effective coupling in the
$\lambda \varphi^{4}$ model.
Here we are dealing with a trivial theory.
If we take the quantum corrections of all scales
into account, the effective coupling in the theory vanishes,
implying that small scale dynamics fundamentally modify
the behavior at larger scales.
Imposing a momentum space cut off in the theory
invalidates triviality and the theory becomes interacting.
A kinetic formalism can be derived
from the closed time path formalism
without resorting to a two-scale interpretation.
However the resulting equations of motion that one obtains
are kinetic equations for the 2 point Green functions,
rather than for a quasi-classical phase-space distribution function.

\par
A kinetic theory for the $\lambda \varphi^{4}$
model was originally derived via the operator
formalism in the mid-seventies \cite{Carr}.
We study the kinetic field theory for the
aforementioned model by reformulating the CTP equations
of motion as a transport theory, and including
up to 4-point initial correlations.
A similar approach can be found in \cite{Cal},
where however only initial two-point correlations
have been taken into account.
In section two, we give a brief
introduction to the CTP formalism
in this context,
derive the transport equations,
and clarify their physical content.
Our view parallels the physical
interpretation put forward in the
early work by Carruthers and Zachariasen
\cite{Carr}.
Initial conditions for these equations
are shown to break a symmetry underlying
{\it any \/} CTP field theory in the third section.
The Ward-Takahashi identities corresponding
to this broken symmetry express direct relationships
between the Green functions and
the initial conditions.
They also represent a set of kinetic equations
generalized to the n-point Green functions.
For the $\lambda \varphi ^{4}$ model,
in which the symmetry group is SO(1,1),
these identities are given.
In particular, the Noether current
corresponding to this symmetry
is interpreted physically.
In addition, it is shown that the
dynamical evolution of the 2-point
functions guarantees
energy-momentum conservation.
In the fourth section we detail this explicitly
by demonstrating that the divergence of the energy-
momentum tensor vanishes
simply by using the transport equations.
This allows us to settle some of the questions
concerning energy-momentum
conservation and the use of gradient-expansions in
both
Vlasov equations,
and equations that include a collision-term.
By studying the off-shell behavior
in the free field equations
and recovering the Casimir-effect from our solutions,
we illustrate the intimate connection between inhomogeneities,
off-shellness and initial conditions.
With this program in mind, we now give a brief exposition
of the CTP formalism and the kinetic field theory
that is based upon it.

\section{Kinetic Field Theory}

\par
\indent
Before we turn to the
$\lambda \varphi^{4}$ model in particular,
we review the general structure
of the CTP formalism and
the generalized effective action.
Following this,
we derive the transport equations
for the 2-point functions
and analyse the physical meaning
of the various terms in the equations
and the variables we use.

\subsection{The Closed Time Path Formalism}

\par
\indent
Here we will briefly recapitulate the main ideas
of the CTP formalism
that are relevant for our calculation.
A full treatment of the CTP formalism
can be found elsewhere \cite{Chin}.
Let $\hat{\varphi}$ denote a collection
of field operators.
For simplicity, we take them to be bosonic.
The system is specified by a Lagrangian density
${\cal L}({\hat{\varphi}})$,
and a density operator $\hat{\rho}$,
that enforces the initial conditions.
\par
Let $\mid \phi_{1} >$ be an eigenstate
of the $\hat{\varphi}$ operator satisfying
\begin{equation}
  \hat{\varphi} > \ \mid \phi_{1} = \phi_{1}(x) \ \mid \phi_{1}> \,
\end{equation}
with the c-number function $\phi_{1}(x)$
being one of its eigenvalues.
Let $ \mid \phi_{n} >$ be a complete set.
Then the functional
\begin{equation}
\rho [ \phi_{i}(x) , \phi_{j}(y)] \ = \ < \phi_{i} \mid \hat{\rho} (t=0) \mid
\phi_{j} >
\label{eq:dens}
\end{equation}
contains all the physical information
stored in the density operator at t=0.
Square brackets denote $\rho$ being a functional rather than a funtion.
If the hamiltonian $\cal{H}$ of the system is not
time dependent then,
in the Heisenberg picture,
neither will the operator $\hat{\rho}$ be.
If we now introduce an additional
interaction with a source $j(\vec{x},t)$
\begin{equation}
{\cal H} \rightarrow {\cal H} + \int d^3 {\vec x}
j({\vec x},t) {\hat \varphi} {(\vec x},t) \ ,
\end{equation}
then the time dependence of $\hat{\rho}$ is given by
\begin{eqnarray}
\hat{\rho}(t)
 & = & [ {\cal{T}} \{ \exp ( -\imath \int_{0}^{t} dt' \int d^{3}\vec{x}'
j(\vec{x'},t') \hat{\varphi} (\vec{x}' ,t' )\ )\}\ ] \nonumber \\
 &   & \times \ \hat{\rho} (t=0) \ \times \nonumber \\
 &   & [ \tilde{\cal{T}} \{ \exp ( +\imath \int_{0}^{t} dt'' \int d^{3}
\vec{x}'' j(\vec{x}'' ,t'') \hat{\varphi} (\vec{x}'',t'' )\ )\}\ ]
\label{eq:densti}
\end{eqnarray}
where $\cal{T}$ is the time ordering operator,
and $\tilde{\cal{T}}$ is the {\it anti } time
ordering operator.
It is a useful bookkeeping device to introduce
indices 1 and 2 for the current and field operators
taken on the time ordered or anti time ordered paths
respectively,
that are given in Eq.(\ref{eq:densti}).
Explicitly on the time ordering path, one has
$\hat{\varphi}^{1}({\vec{x}},t)$ and $j^{1}$,
and $\hat{\varphi}^{2}({\vec{x}},t)$ and $j^{2}$
on the anti time ordering path.
In a contravariant notation,
the general indexed n-point Green functions
are defined by
\begin{equation}
{\cal{G}}^{a_{1}...a_{n}} (x_{1},...,x_{n}) =  \langle {\cal{T}}_{p} \{
\hat{\varphi}^{a_{1}}(x_{1}) ... \hat{\varphi}^{a_{n}}(x_{n}) \} \rangle
\label{eq:discdef}
\end{equation}
where ${\cal{T}}_{p}$ orders the field-operators
according to the rule
\begin{equation}
{\cal{T}}_{p} \{ \prod_{n} \ {\hat{\varphi}}^{a_{n}}(x_{n}) \ \}
= \tilde{{\cal{T}}} \{ \prod_{a_{n} = 2} {\hat{\varphi}}^{a_{n}}(x_{n}) \}
 {\cal{T}} \{ \prod_{a_{k} = 1} {\hat{\varphi}}^{a_{k}}(x_{k}) \} \ .
\end{equation}
\par
The action ${\cal{S}} [\phi ] = \int d^{4} x \  {\cal{L}}$
enters in the functional integral representation
of the trace of the density matrix,
that is obtained by writing the time-evolution operators
in Eq.(\ref{eq:densti}) as path-integrals,
and using the functional Eq.(\ref{eq:dens})
\begin{eqnarray}
\label{eq:funcint}
{\cal{Z}}[ j^{a} , \rho] & = & Tr \{ \hat{\rho}(t= \infty) \} \nonumber \\
  & = & \int D\phi^{1} \ D \phi^{2} \ \exp \ \imath \{ {\cal{S}} [ \phi^{1} ] +
  j^{1} \circ \phi^{1} - {\cal{S}} [ \phi^{2} ] - j^{2} \circ \phi^{2} \ \}
  \rho [ \phi^{1} , \phi^{2} ] \ ,
\end{eqnarray}
using the abbreviation {$j \circ \phi = \int d^{4}x j(x) \phi(x) $}.
\par
It is convenient to replace the fields $\phi^{1}$ and $\phi^{2}$
by the doublet {$\varphi^{a} = ( \phi^{1} , \phi^{2} )$},
and introduce the metric
\begin{equation}
\label{eq:metric}
  c_{ab} = c^{ab} = \left( \begin{array}{cc} 1 & 0 \\ 0 & -1 \end{array}
\right)
\end{equation}
on this internal space.
For example, the source term
can then be written as
\begin{equation}
  j_{1} \circ \phi_{1} - j_{2} \circ \phi_{2} = c_{ab} j^{a} \circ \varphi^{b}
  = j_{a} \circ \varphi^{a}
\end{equation}
by contracting the fields $\phi$ with the metric $c^{ab}$.
We will refer to these indices $a,b$ as CTP indices.
We define a CTP action $S_{CTP} = S[\phi_1] - S[\phi_{2}] $
and write for the partition function $\cal{Z}$
\begin{equation}
{\cal{Z}} [j_{a}, \rho] = \int D \varphi^{b} \exp \ \imath \{ S_{CTP} [
\varphi^{b}]
+ j_{a} \circ \varphi^{a} \ \} \rho [\varphi^{b}] \ .
\end{equation}
The general Green functions ${\cal{G}}^{abc...}$
are com\-puted now by ta\-king deri\-vatives
of Z with res\-pect to the source $j_{a}(x)$.
\par
By expanding the density functional in Eq.(\ref{eq:dens}),
in powers of the field {$\varphi^{a}$}
\begin{equation}
  \rho [\varphi^{a}] = \exp \ \imath \{ \sum_{n=0}^{\infty} \frac{1}{n!}
  K_{a_{1}...a_{n}} \circ \varphi^{a_{1}} \circ ... \circ \varphi^{a_{n}} \ \}
\ ,
\end{equation}
one obtains the expression for the generating functional
of the general n-point Green functions
$ {\cal{G}}^{a_{1} ... a_{n}}$  to be,
\begin{equation}
\label{eq:genfunc}
  {\cal{Z}} [j_{a}, K_{ab}, ...] = \int D {\varphi^{b}} \exp \ \imath
\{ S_{CTP} [ {\varphi^{b}} ] + {j_{a} \circ \varphi^{a}} + {\sum_{n=0}^{\infty}
\frac{1}{n!}}
{K_{a_{0}...a_{n}} \circ \varphi^{a_{0}} \circ ... \circ \varphi_{a_{n}} }\} .
\end{equation}
{}From Eq.(\ref{eq:genfunc}), one sees that
the non-local sources $K_{abc...}$ act as
sources for the full Green functions
${\cal{G}}^{abc...}$.
In this way they prescribe the initial values
of these Green functions.
In a diagrammatic expansion of
Eq.(\ref{eq:genfunc}),
the non-vanishing K's
are part of the tree level contributions
to the non-equilibrium vertex functions,
although they contain contributions from
all loop orders.

\par
The generating functional \ef \
of the connected Green functions
$G^{abc...}(x,y,z,...)$ is the logarithm
of $ \cal{Z} $
\begin{equation}
  \ef[j_{a}, K_{ab}, ...] = - \imath
  \log \{ {{\cal{Z}}} [ j_{a} , K_{ab} , ...] \} \ .
\end{equation}
and the connected Green functions can be derived from
this in the stan\-dard fashion.
An alternative method of obtaining the connected Green
fuctions is as follows.
One derives the {\it full \/}
Green func\-tions ${\cal{G}}_{abc...}$ via
dif\-ferentia\-tion of \ef \ with res\-pect to
the non-\-local sources
\begin{equation}
\label{eq:gdef}
{\cal{G}}^{a_{1},...,a_{n}} (x_{1},...,x_{n}) =
  n! \frac{ \delta \ef }{ \delta K_{a{1} ... a_{n} }}
  [j_{b}, K_{bc}, ... ] \ ,
\end{equation}
and obtains the connected Green functions by
subtracting out the disconnected contributions.
This procedure is most simply convenient in
deri\-ving the hierarchy of coupled equa\-tions
of mo\-tion for the (connected) Green func\-tions
in a direct fashion.
In order to obtain these equations for the $G^{abc..}$,
and $G^{ab}$ in particular,
we construct an apropriate action
functional by means of a Legendre
transform with respect to the non-local sources.
Let us denote the mean-field $G^{a}(x)$ as \mfi{a}.
Defining \gea as the Legendre
transform of \ef,
\begin{equation}
\label{geacdef}
  \Gamma [\mfi{a}(x), G^{ab}(x,y), ...]
= \ef [ j_{a},K_{ab},...] - j_{a} \circ \mfi{a} -
  \frac{1}{2} K_{ab} \circ
  \{ G^{ab} + \mfi{a} \mfi{b} \} - \ldots
\end{equation}
the mean-field \mfi{a} and the con\-nected
2-point and n-point fun\-ctions satis\-fy
the equa\-tions
\begin{eqnarray}
  \frac{ \delta \gea}{\delta \mfi{a}} [ \mfi{c} ,G^{cd},...] & = &
- j_{a} -  K_{ab} \circ \mfi{b} - ... \\
  \frac{ \delta \gea}{ \delta G^{ab}} [ \mfi{c} ,G^{cd},...] & = &
- \sum_{n=2}^{\infty} \frac{1}{n!} K_{a_{1}...a_{n} } \circ \frac{ \delta }{
\delta G^{ab } }{\cal{G} }^{a_{1}...a_{n} } \\
... & ... & ... \nonumber \\
  \frac{ \delta \gea}{ \delta G^{b_{1}...b_{m} } } [ \mfi{c} ,G^{cd},...] & = &
- \sum_{n=m}^{\infty} \frac{1}{n!} K_{a_{1}...a_{n} } \circ \frac{ \delta }{
\delta G^{b_{1}...b_{m} } }{\cal{G} }^{a_{1}...a_{n} } \nonumber \\ .
\end{eqnarray}
The general Green functions ${\cal{G}}^{abc...}$
can be expressed in terms of the connected Green
functions $G^{abc...}$ yielding the formal equations
of motion. They are formal in the sense that \gea
\ has not been computed explicitly.
For our purpose
this will not be neccesary.
The physical implications
of such a Legendre transform were treated,
for example, in \cite{Dan} and \cite{Schmiwi}.
At this point we leave the formal development
of the CTP-formalism and turn
to its application to the $ \lambda \varphi^{4}$ model.
\par
We consider the Lagrangian density
\begin{equation}
{\cal{L}} = \frac{1}{2} \partial_{\mu}\varphi \partial^{\mu}\varphi
- \frac{m^2}{2} \varphi^{2} - \frac{\lambda}{4!}\varphi^{4} \ .
\end{equation}
The corresponding CTP action is then
\begin{equation}
S_{CTP} [\varphi^{a}] = \int d^{4}x \ \frac{1}{2} c_{ab} \{
\partial_{\mu} \varphi^{a} \partial^{\mu} \varphi^{b} - m^{2}
\varphi^{a} \varphi^{b} \}
- \frac{\lambda}{4!} h_{abcd}\varphi^{a}\varphi^{b}\varphi^{c}\varphi^{d}
\end{equation}
where we have defined
\begin{equation}
h_{abcd} = \left\{
\begin{array}{ll}
 1 & \mbox{when $a=b=c=d=1$} \\
-1 & \mbox{when $a=b=c=d=2$} \\
\end{array}
\right\} \ .
\end{equation}
In the next subsection we focus on the equations of motion
for the 2-point connected Green functions,
since they lie at the heart
of the kinetic formulation of the non-equilibrium theory.

\subsection{The Kinetic Formulation}

\par
\indent
In this sub\-section we de\-rive
the equa\-tions of mo\-tion specifically
for the 2-point fun\-ctions $G^{ab}(x,y)$
and write them as kine\-tic equa\-tions.
In particular, a hierarchy of coupled equations
for the $G^{abc...}$ is obtained in which the
two-point function $G^{ab}$ is embedded.
For prac\-ti\-cal cal\-cu\-la\-tions,
this hier\-archy of equa\-tions for the $G^{abc...}$
must be truncated.
Calzetta and Hu \cite{Cal},
Kandrup and Hu \cite{Kan} and,
in a ge\-neral con\-text,
Hu \cite{Hu},
have argued that this in\-tro\-du\-ces
dis\-si\-pa\-tive phe\-no\-me\-na in the model.
On the other hand,
if the field is ini\-tial\-ly free
and the den\-sity opera\-tor is diagonal in
momentum space,
an ex\-pli\-cit cal\-cu\-la\-tion
of the non local sour\-ces K is feasable,
and shows that the series trun\-cates
af\-ter the two point source.
Consequently if one is wil\-ling
to accept li\-mi\-ta\-tions
on the range of initial conditions,
the series of the K's can be trun\-cated.
It is in this spirit that we will trun\-cate the
expan\-sion
of ${\rho [\varphi^{a}]}$ after $K_{abcd}(u,v,w,x)$.

\subsubsection{The Effective Action for the $\lambda \varphi^{4}$ Model}

\par
\indent
Truncating \gea \ after the fourth term,
leads to the explicit form
\begin{eqnarray}
\gea [\mfi{a} , G^{ab}, G^{abc}, G^{abcd} ] & = & \ef [j_{a} , K_{ab} , K_{abc}
, K_{abcd}]
\\ - j_{a} \circ \mfi{a} -  \frac{1}{2}K_{ab} \circ {\cal{G}}^{ab} &
- \frac{1}{3!}K_{abc} \circ {\cal{G}}^{abc} - & \frac{1}{4!} K_{abcd}
\circ {\cal{G}}^{abcd} \nonumber
\end{eqnarray}
implying that \mfi{a} ,
${\cal{G}}^{ab}(x,y)$,
${\cal{G}}^{abc}(x,y,z)$ and ${\cal{G}}^{abcd}(w,x,y,z)$
can be chosen as in\-depen\-dent variables.
A more con\-venient choice are
the 2-point {\it connected} Green func\-tions $G^{ac}(x,y)$
\begin{equation}
G^{ab} (x,y) = 2 {\cal{G}}^{ab}(x,y) - \mfi{a}(x) \mfi{b}(y),
\end{equation}
and the 3- and 4- ir\-redu\-cible
ver\-tices of the theory,
$\alpha^{3}_{abc}(x,y,z)$ and
$\alpha^{4}_{abcd}(w,x,y,z)$ respectively,
defined by
\begin{equation}
\imath \alpha^{3}_{abc} \circ G^{aa'} G^{bb'} G^{cc'}
= 3! {\cal{G}}^{a' b' c'} - G^{a' b'}\mfi{c'}
- G^{b' c'}\mfi{a'} - G^{c' a'}\mfi{b'} -
\mfi{a'} \mfi{b'} \mfi{c'}
\end{equation}
and,
\begin{eqnarray}
\imath \alpha^{4}_{abcd} \circ G^{aa'} G^{bb'} G^{cc'} G^{dd'} &  +  & G^{aa'}
G^{bb'} \circ
\imath \alpha^{3}_{abm} \circ G^{mm'}
\circ \imath \alpha^{3}_{m' cd} \circ G^{cc'} G^{dd'} +  permutations \nonumber
\\
& = & 4! {\cal{G}}^{a' b' c' d'} - disconnected \ terms \ ,
\end{eqnarray}
and which are re\-pre\-sented dia\-gram\-matical\-ly
in fig.1 \ .
As shown by Weldon \cite{Weld},
in the non-equili\-brium formulation these vertices
are related to the equili\-bration rate of the dis\-tribu\-tion
of quasi-par\-ticles over the available energy-\-momen\-tum space,
and not to the decay-\-rate for a single par\-ticle.

\subsubsection{The Equations of Motion}

\par
\indent
It is possible to extract all
information needed to construct the
equations of motion for $G^{ab}$ from the
dynamics of the mean field \mfi{a} alone.
For the sake of convenience we will
always assume that the sources $K_{abc}$
and $K_{abcd}$ include the tree-level
vertices.
We proceed as follows.
The equation of motion for \mfi{a}
is found by solving the identity
\begin{equation}
0 = \int D {\varphi^{b}} \ \frac{\delta}{\delta \varphi^{a}} \exp \ \imath
\{ S_{0 \ CTP} [ {\varphi^{b}} ] + {j_{a} \circ \varphi^{a}} + {\sum_{n=2}^{4}
\frac{1}{n!}} {K_{a_{0}...a_{n}} \circ \varphi^{a_{0}} \circ
... \circ \varphi^{a_{n}} }\} ,
\end{equation}
where the free CTP action \ $S_{0 \ CTP}$,
given by
\begin{equation}
 S_{0 \ CTP} = \int d^{4}x \frac{c_{ab}}{2} \{ \partial_{\nu}\varphi^{a}
 \partial^{\nu}\varphi^{b} + m^{2}\varphi^{a}\varphi^{b} \}  ,
\end{equation}
appears due to our choice of absorbing
the interaction vertices in the K's.
Working this out yields
\begin{equation}
\label{eq:id}
G_{(0) \ ab }^{-1} \circ \mfi{b} =  j_{a}  + \sum_{n=2}^{4} \frac{1}{n!} K_{a
a_{2}...a_{n}} \circ {\cal{G}}^{a_{2} ...a_{n} } ,
\end{equation}
where we have denoted as $ G_{(0) \ ab }^{-1}$
the inverse of the free propagator
\begin{equation}
G_{(0) \ ab }^{-1} = c_{ab} \{ {\Box}^{2} - m^{2} \} \ .
\end{equation}
{}From  Eq.[\ref{geacdef}],
it follows formally that
\begin{equation}
\label{eq:eqm1}
\frac{\delta \gea}{\delta \mfi{a}} = - j_{a} - K_{ab} \circ \mfi{b}
- \frac{1}{2} K_{abc} \circ \frac{\delta {\cal{G}}_{bcd} }{\delta \mfi{d} }
-\frac{1}{3!}K_{abcd} \circ \frac{\delta {\cal{G}}_{bcde} }{\delta \mfi{e} } \
{}.
\end{equation}
Comparing Eq.[\ref{eq:id}] and Eq.[\ref{eq:eqm1}],
and using
\begin{equation}
\frac{\delta {\cal{G}}^{a a_{2}...a_{n}} }{\delta \mfi{a}} =
n {\cal{G}}^{a_{2}...a_{n}}
\end{equation}
we can identify
the l.h.s. of Eq.[\ref{eq:eqm1}] to be
\begin{equation}
\label{eq:prop}
\frac{\delta \gea}{\delta \mfi{a}} = G_{(0) \ ab }^{-1} [G^{ab}] \circ \mfi{b}
\ .
\end{equation}
$G_{(0) \ ab}^{-1}$ is now a func\-tional
of the full pro\-paga\-tor.
The ex\-plicit con\-nec\-tion is given by
the usual Dyson equation.
As in the zero-temperature formalism,
Eq.[\ref{eq:prop}] can be formally integrated
with respect to \mfi{a} yielding
\begin{equation}
\gea [\mfi{a} , G^{ab}, \alpha_{abc}^{3} , \alpha_{abcd}^{4} ]
= \gea_{0} [\mfi{a} , G^{ab}] + \gea_{int} [G^{ab} , \alpha_{abc}^{3} ,
\alpha_{abcd}^{4} ] \ .
\end{equation}
$\gea_{int}$  can now be constructed from
vacuum diagrams alone, in terms of $G^{ab}$,
$\alpha_{abc}^{3}$ and $\alpha_{abcd}^{4}$ ,
which simplifies the calculation considerably.
To see this,
con\-si\-der,
for example,
a dia\-gram D
per\-taining to $\gea_{int}$
with $n_{G}$ pro\-pa\-ga\-tors,
$n_{3}$ 3-point ver\-ti\-ces and
$n_{4}$ 4-point ver\-ti\-ces.
It has no ex\-ter\-nal lines,
since they would end in an \mfi{a},
so
\begin{equation}
\label{eq:comb}
n_{G} = \frac{1}{2} \{ 3 n_{3} + 4 n_{4} \} \ .
\end{equation}
Let $\gea_{int}(D)$ be the Feynman-amplitude
of the diagram D contributing to the
functional $\gea_{int}$.
Cutting through one propagator line
in D represents the action
of a functional derivative,
with respect to $G^{ab}$,
on $\gea_{int}(D)$.
The operator $G^{ab} \circ
\frac{\delta}{\delta G^{ab}}$
thus counts the number of
propagators in every diagram.
Using the appropriate counting operators
for the vertices, and using the fact that
Eq.[\ref{eq:comb}] holds for all
diagrams contributing to $\gea_{int}$,
one finds
\begin{equation}
G^{ab} \circ \frac{\delta }{\delta G^{ab}} \gea_{int}
= \frac{3}{2} \alpha_{abc}^{3} \circ \frac{\delta}{\delta \alpha_{abc}^{3}}
\gea_{int} + 2 \alpha_{abcd}^{4} \circ \frac{\delta}{\delta \alpha_{abcd}^{4}}
\gea_{int} \ .
\end{equation}
The formal equation of motion for $G^{ab}$,
after truncating the series of the $K_{abc..}$,
is
\begin{equation}
\label{eq:eqm2}
  \frac{ \delta \gea}{ \delta G^{ef}} [ \mfi{c} ,G^{cd}, \alpha_{abc}^{3},
\alpha_{abcd}^{4} ]  =
- \frac{1}{2} K_{ef} - \frac{1}{3!} K_{abc} \circ \frac{\delta
{\cal{G}}_{abc} }{ \delta G^{ef} } - \frac{1}{4!} K_{abcd} \circ \frac{\delta
{\cal{G}}_{abcd} }{ \delta G^{ef} } \ .
\end{equation}
The l.h.s. of this equation can be expressed as
\begin{equation}
\frac{ \delta \gea}{ \delta G^{ef}} [ \mfi{c} ,G^{cd}, \alpha_{abc}^{3},
\alpha_{abcd}^{4} ]  =
\frac{ \delta \gea_{0}}{ \delta G^{ef}} [ \mfi{c} ,G^{cd}, \alpha_{abc}^{3},
\alpha_{abcd}^{4} ] +
\frac{ \delta \gea_{int}}{ \delta G^{ef}} [ \mfi{c} ,G^{cd}, \alpha_{abc}^{3},
\alpha_{abcd}^{4} ].
\end{equation}
The first term here can be explicitly calculated
from the functional integral for ${\cal{Z}}$
\begin{equation}
\gea_{0} [\mfi{a},G^{ab},\alpha_{abc}^{3},\alpha_{abcd}^{4}] =
S_{o \ CTP}[\mfi{a}] - \frac{\imath}{2}\log \{G^{ab}\}
+ \frac{\imath}{2} G_{(0) \ ab}^{-1} \circ G^{ab} .
\end{equation}
The last term can be written out
using the formal equa\-tions of mo\-tion
of the ver\-ti\-ces $\alpha_{abc}^3$
\begin{equation}
\frac{\delta \gea}{\delta \alpha_{abc}^3} =
- \frac{\imath}{3!} G^{ad}G^{be}G^{cf} \circ \{ K_{def} - K_{defg}\mfi{g} \}
- \frac{1}{4} G^{ad}G^{be} \circ K_{defg} \circ G^{fh}G^{gi} \circ
\alpha_{hij}^{3} \circ G^{jc} ,
\end{equation}
and $\alpha_{abcd}^4$,
\begin{equation}
\frac{\delta \gea}{\delta \alpha_{abcd}^4} =
\frac{\imath}{4!} G^{ae}G^{bf}G^{cg}G^{dh} \circ K_{efgh} ,
\end{equation}
obtained from Eq.[\ref{geacdef}].
Af\-ter some tedious al\-ge\-bra,
we find the equation for $G^{ab}$ to be
\begin{eqnarray}
\label{eq:eqmo}
\imath G_{ab}^{-1} = &
\imath G_{(0) \  ab}^{-1} &
- \ \frac{\lambda}{2}h_{abcd}
\circ \{ \mfi{c} \mfi{d} + G^{cd} \} \ \nonumber \\
     &
- \frac{\imath \lambda}{3!}h_{becd} \ \circ &
\alpha_{afgh}^{4} \circ G^{fc}
 G^{gd} G^{he} \nonumber \\
     &
- \frac{\imath \lambda}{2}h_{befg} \ \circ &
\mfi{g} \alpha_{acd}^{3} \circ G^{ce} G^{df} \\
     &
-\frac{\lambda}{4}h_{bcde} \  \circ &
\alpha_{afm}^{3} \alpha_{lgh}^{3} \circ G^{fc} G^{ml} G^{gd} G^{he} \nonumber
\\
     &
-\frac{\lambda}{8}h_{clde} \circ &
\alpha_{fma}^{3} \alpha_{bgh}^{3} \circ G^{fc} G^{ml} G^{gd} G^{he} \nonumber
\end{eqnarray}
The solution of this equation
requires knowledge of both vertices
$\alpha_{abc}^{3}$ and $\alpha_{abcd}^{4}$.
It describes the exact 2-point Green function
for the non-equilibrium system,
with initial conditions constrained by
the truncation of the K's after
$K_{abcd}$.
At this point, it makes no explicit reference
to the dynamical role of inhomogeneities
and off-shellness.
The kinetic representation of
Eq.[\ref{eq:eqmo}]
however allows one to address
these questions directly.
Since we aim at understanding
the simplest effects of inhomogeneities
in the field theoretic setting,
there is no apparent need for allowing
spontaneous symmetry breaking.
This allows one to make an appreciable
technical simplification of Eq.[\ref{eq:eqmo}].
We thus set the mean-field and
the 3-vertex to zero, so that
the last three terms vanish.
\par
We progress toward the kinetic
representation of the non-equilibrium dynamics
by multiplying Eq.(\ref{eq:eqmo}) once from the right
by $G^{bc}$ and contracting over the index $b$ yielding
\begin{equation}
\label{eq:inv1}
\imath \delta_{a}^{\ c} = \imath G_{(0) \ ab}^{-1} \circ G^{bc}
 - \frac{\lambda}{2} h_{abef} \circ G^{ed} G^{bc} -
 \frac{\imath \lambda}{3!} h_{beld} \circ  \alpha_{afgh} \circ
 G^{fl} G^{gd} G^{he} G^{bc} .
\end{equation}
Here the differential operator
$G_{(0) \ ab}^{-1}$ acts on $G^{ab}$.
Multiplication of Eq.(\ref{eq:eqmo})
from the left by $G^{ca}$
and contracting over the index $a$ yields
\begin{equation}
\label{eq:inv2}
\imath \delta_{\ b}^{a}  = \imath G^{ca} \circ G_{(0) \ ab}^{-1}
 - \frac{\lambda}{2} h_{abef} \circ G^{ca} G^{ef} - \frac{\imath
 \lambda}{3!} G^{ca} h_{beld} \circ \alpha_{afgh} \circ
 G^{fl} G^{gd} G^{he} .
\end{equation}
In doing so we have formally inverted
the operator $G_{ab}^{-1}(x,y)$.
In order to do so properly it
is neccesary to specify its kernel,
i.e. the set of functions $f(x)$ for which
\begin{eqnarray}
G_{ab}^{-1} (x,y) \circ f(y,z) & = & 0 \nonumber \\
G_{ab}^{-1} (x,y) \circ f(x,z) & = & 0  .
\end{eqnarray}
This coincides with specifying boundary
conditions on Eq.(\ref{eq:inv1})
and Eq.(\ref{eq:inv2}).
Allowing inhomogeneities in the
boundary conditions entails the need
for both Eq.(\ref{eq:inv1}) and
Eq.(\ref{eq:inv2}) instead of simply
Eq.(\ref{eq:eqmo}).
\par
For a homo\-geneous sys\-tem ,
$G^{ab}(x,y)$
de\-pends only on
the re\-lative coordinate
${\rx = x - y}$.
It is there\-fore na\-tu\-ral
to con\-sider $G^{ab}$ as
a func\-tion of $\rx$ and
the cen\-tre of mass coor\-dinate
${X = \frac{1}{2} \{ x + y \}}$.
Its 4-Fourier trans\-form
$G^{ab}(P,\rp )$
\begin{equation}
G^{ab}(P,\rp) = \int d^{4}X d^{4}\rx \ \ e^{\imath PX + \imath \rx \rp}
G^{ab}(X,\rx )
\end{equation}
is related to the Wigner transform $G^{ab}(X,\rp )$ by
\begin{equation}
G^{ab}(X,\rp ) = \int \frac{d^{4} P}{(2 \pi)^{4}} e^{- \imath PX}
G^{ab}(P,\rp ) \ .
\end{equation}
Then Eq.(\ref{eq:inv1}) and
Eq.(\ref{eq:inv2})
in terms of $G^{ab}(P,\rp )$ become
\begin{eqnarray}
\imath (\rp + \frac{P}{2} )^{2} G^{ab}(P,\rp )
& - & \frac{\lambda}{2} \int \frac{d^{4}P' }{(2 \pi )^{4}}
\frac{d^{4} \rp '}{(2 \pi )^{4}} G^{aa}(P',\rp ')\exp
\{ \frac{1}{2} P'. \partial_{\rp } \} G^{ab}(P-P',\rp ) \nonumber \\
& = & \frac{\imath \lambda}{12} \int\frac{d^{4} P'}{(2 \pi )^{4}}
\Sigma_{\ c}^{a}(P-P',\rp ) \Delta^{-} G^{cb}(P',\rp ) , \nonumber \\
\end{eqnarray}
and
\begin{eqnarray}
\imath (\rp - \frac{P}{2} )^{2} G^{ab}(P,\rp )
& - & \frac{\lambda}{2} \int \frac{d^{4}P' }{(2 \pi )^{4}}
\frac{d^{4} \rp '}{(2 \pi )^{4}} G^{aa}(P',\rp ')\exp
\{ - \frac{1}{2} P'. \partial_{\rp } \} G^{ab}(P-P',\rp ) \nonumber \\
& = & \frac{\imath \lambda}{12} \int\frac{d^{4} P'}{(2 \pi )^{4}}
\Sigma_{\ c}^{a}(P-P',\rp ) \Delta^{+} G^{cb}(P',\rp ) , \nonumber \\
\end{eqnarray}
where the 2-loop self-energy,
$\Sigma_{ab}$,
is given by
\begin{equation}
\label{eq:se}
\Sigma_{ab} = h_{bcde} \circ \alpha_{ac' d' e'}^{4}
\circ G^{cc'} G^{dd'} G^{ee'} ,
\end{equation}
and we have introduced
the gradient operators
\begin{equation}
\label{eq:de}
\Delta^{\pm} = \exp \{ \pm \frac{1}{2} \{ (P-P') .\partial_{\rp}^{\rightarrow}
- P'.\partial_{\rp}^{\leftarrow} \} \ .
\end{equation}
In this expression, the ar\-rows in\-di\-ca\-te the
di\-rec\-tion in which the
dif\-feren\-tia\-tion acts.
By subtracting Eq.(\ref{eq:inv2})
from Eq.(\ref{eq:inv1}) we obtain
the kinetic equation
\begin{eqnarray}
\label{eq:ke}
\imath \rp . P G^{ab}(P,\rp  )
& - & \frac{\lambda}{2} \int \frac{d^{4}P' }{(2 \pi )^{4}} \frac{d^{4} \rp
'}{(2 \pi )^{4}}
 G^{aa}(P',\rp ')\sinh \{ \frac{1}{2} P'. \partial_{\rp } \} G^{ab}(P-P',\rp )
\nonumber \\
& = & \frac{\imath \lambda}{12} \int\frac{d^{4} P'}{(2 \pi )^{4}}
\{ \Sigma_{\ c}^{a} \Delta^{-} G^{cb} - \Sigma_{c}^{\ b} \Delta^{+} G^{ac}
\}\nonumber \\
\end{eqnarray}
and,
by averaging the equations,
Eq.(\ref{eq:inv2}) and Eq.(\ref{eq:inv1}),
the constraint equation
\begin{eqnarray}
\label{eq:ce}
\imath \{ {\rp}^{2} + \frac{1}{4} P^{2}- m^{2} \} G^{ab}(P,\rp )
& - & \frac{\lambda}{2} \int \frac{d^{4}P'}{(2 \pi )^{4}} \frac{d^{4}\rp '}{(2
\pi )^{4}}
G^{aa}(P',\rp ')\cosh \{ \frac{1}{2} P' . \partial_{\rp } \} G^{ab}(P-P',\rp )
\nonumber \\
& - & \frac{\imath \lambda}{12} \int\frac{d^{4} P'}{(2 \pi )^{4}}
\{ \Sigma_{\ c}^{a} \Delta^{-} G^{cb} + \Sigma_{c}^{\ b} \Delta^{+} G^{ac} \}
\nonumber \\
& = & \imath c^{ab} (2 \pi)^{4} \delta^{4}(P) \ , \nonumber \\
\end{eqnarray}
is obtained.
The so-called gradient expansion for the
Green functions can now easily be obtained by
a truncation of the series expansions
of the operators in Eq.[\ref{eq:de}]
or alternatively the sinh and cosh
occuring in the Eqs.[\ref{eq:ke}]
and [\ref{eq:ce}].
It is a well-\-defined ap\-proxima\-tion whenever
$ \mid P'_{\nu}\partial_{\rp }^{\nu}\mid$ is small.
Since $P'$ is integrated over,
this suggests that
the $P'$-integration is
regulated at an ultraviolet limit
$\Lambda$ and the gradient is restricted
similarly, i.e. $\mid \partial_{\rp} \mid < \Lambda^{-1}$.
Such a cutoff neccesarily in\-tro\-du\-ces a
length scale.
Furthermore it also alters the nature
of the $\lambda \varphi^{4}$ model
substantially.
Both topics have been discussed in the introduction.
If the opera\-tors are to act on
func\-tions con\-tai\-ning re\-sonance-like
sin\-gularities in $\rp$,
then the ap\-proxi\-mation
is no longer valid close to
the pole for any value of $P$.
\par
We close this section with a discussion
of the physical content of the various terms
in the kinetic and constraint equations.

\subsubsection{Physical Interpretation}

\par
\indent
Under homogeneous boundary conditions,
such as global thermal equilibrium (T.E.),
we have
\begin{equation}
G^{ab}(P,\rp ) \propto \delta^{4} (P) \ ,
\end{equation}
which,
when substituted into Eq.(\ref{eq:ke}),
makes the kinetic equation void.
The constraint equation reduces
to the Dyson equation.
The variable $\rp$ can then be identified
with the momentum of the propagating boson,
and, for example,
$G^{21}(\rp )$ can be written
in terms of a Bose-Einstein
distribution function $f(\rp )$ as
\begin{equation}
\label{eq:G}
G^{21}(\rp ) = \{ \theta(\rp^{0}) + f(\rp ) \} \delta ( \rp^{2} - m^{2} ) \ .
\end{equation}
Similarly, one can identify
$\rp . P \ G^{ab}(P,\rp )$ as the
flow-term of the kinetic equation.
The terms containing the sinh function in
Eq.(\ref{eq:ke}) describe the ef\-fect of
the inter\-actions with a back\-ground of
spon\-taneously created (off-shell)
par\-ticles.
Due to the analogy with the Vlasov
interaction in plasma physics,
we refer to it as the Vlasov term.
The Vlasov term of Eq.(\ref{eq:ke})
describes a strictly local interaction
of a single particle with the
density of other particles.
There is no long range component to this.
This situation may change however
when a mean field $\mfi{a}$ is
present that contains also non-local
information.
\par
The term in the r.h.s. of the kinetic equa\-tion
is the col\-li\-sion term.
On-shell, it des\-cribes the ef\-fect
of two-particle inter\-actions
on the one-\-par\-ticle
dis\-tri\-bu\-tion function $f$.
The relation between the
two-loop self-energy on the
one hand and the decay of the bosons
is found from the cutting equations
implementing unitarity \cite{Velt}.
The Boltzmann equation for a Bose-gas
is obtained from the kinetic equation
by inserting the form of the $G^{ab}$
from Eq.[\ref{eq:G}] into Eq.(\ref{eq:ke}),
generalizing $f(\rp )$ to $f(P,\rp )$ and
neglecting all gradients in \rp .
This equation then describes on-shell bosons.
The previous remarks on the validity of
the gradient expansion apply here.
On-shellness is {\it not \/} compatible
with neglecting gradients.
The problem can be softened by asssuming
that the boson self-energy has an
imaginary part,
so that \cite{Weld} the distribution $f$
is not in equilibrium.
In that case, the $\delta (\rp^{2} - m^{2})$
would be replaced by a Lorentzian \cite{Mal},
thus allowing $\mid P'.\partial_{\rp} \mid$
to stay bounded for finite $P'$.
Such an imaginary parts does not arise
as long as the collision term is neglected.
\par
The physical meaning of the
variable $P$ can be clarified
by analyzing the transition amplitude
between two states,
$ <1, p^{\mu} \mid$ and $\mid q^{\mu},2>$,
that contain at least
one, reducible,
on-shell particle with
initial incoming momentum $p$,
and final outgoing momentum $q$.
In our convention the momenta are
directed towards the
interaction vertex.
By definition
\begin{equation}
p^{\mu} - \{ - q^{\mu} \} = P^{\mu} .
\end{equation}
So $P$ is the momentum loss due to
interactions with a non-trivial background.
The fact that for homo\-geneous sys\-tems
$P=0$ indicates that the inter\-pretation is of
a statis\-tical na\-ture.
In homo\-geneous sys\-tems the particles will
inter\-act with the back\-ground,
but the effects will average out.
It is easy to show that
\begin{equation}
p^2 = q^2 = \rp^2 = m^2 \Longrightarrow p^{\nu} = q^{\nu} \ .
\end{equation}
If $G^{ab}(P,\rp )$ is restricted to
being on-shell,
with $\rp$ on-shell as well,
then it contains no dynamical
information on the inhomogeneities.
\par
Thus far, we have derived the kinetic field theory
for self interacting bosons.
In the homogeneous limit the
functions $G^{ab}(P,\rp)$ describe the
distribution of on-shell bosons
over the available energy and
momentum states.
In this limit, $\rp$ is the
momentum of the propagating bosons.
When we only require the asymptotic
states to be on-shell
$P$ can be interpreted as the loss
of momentum due to interactions
with inhomogeneities in the system.
\par
In the next section, we will discuss
a symmetry property of non-equilibrium
field theories.
It generates identities among Green
functions that, among other things,
express the impact of initial conditions
on these Green functions.
It serves to yield a satis\-fying inter\-pretation
of the 2-point functions and their arguments
beyond the limit of quasi-homogeneity.

\newpage

\section{The Broken SO(1,1) Symmetry of
Non-Equilibrium Field Theory}

\subsection{SO(1,1) Symmetry}

\par
\indent
In this section, we focus on a symmetry
underlying the non-equilibrium dynamics
of the scalar field.
But first let us make some general remarks.
The transition from standard quantum field theory
to the CTP formalism gives rise to a doubling
of the number of fields.
Every physical field becomes a CTP doublet,
and on the internal space of these CTP doublets,
we have defined the degenerate metric $c_{ab}$.
Now consider a theory containing an n-tuple of
fermions (and/or bosons) symmetric under
SU(n) transformations.
The free CTP action will then,
through doubling,
be symmetric under SU(n,n).
As expected, this symmetry is broken,
although an $SU(n) \bigotimes SU(n)$
subgroup remains.
Now SU(n,n) has $(2n)^{2} - 1 $ generators,
whereas $SU(n) \bigotimes SU(n)$ only has
$2n^{2} - 2$ generators.
Consequently the Noether theorem will
yield $2n^{2} + 1$ currents whose
non-conservation can be computed exactly.
For a broken symmetry, Ward identities can
be derived and considered useful if,
either the symmetry breaking terms are
small or, symmetry violation is calculable.
Here we will consider the symmetry
relevant to the $\lambda \varphi^{4}$
model.
The real scalar field has a trivial SO(1)
symmetry, which under CTP doubling becomes
a broken SO(1,1) symmetry.
To our know\-ledge this has gone
un\-no\-ticed, although there is a relation to
the SU(1,1) lie algebra used in quantum optics
\cite{Ban}.
This section focus\-ses on the SO(1,1)
sym\-metry of the free real scalar field.
We will give the Noe\-ther cur\-rent,
the Ward-\-Takahashi iden\-tities,
dis\-cuss its breaking by
inter\-actions and initial
con\-ditions.

\par
The CTP action for the free field
\begin{equation}
S_{CTP} [\varphi^{a}] = \int d^{4}x \ \frac{1}{2} c_{ab} \{
\partial_{\mu} \varphi^{a} \partial^{mu} \varphi^{b} - m^{2}
\varphi^{a} \varphi^{b} \}
\end{equation}
is in\-variant under glo\-bal SO(1,1)
trans\-for\-mations.
The generator of these is
\begin{equation}
\tau_{a}^{b} = \left(
\begin{array}{cr}
0 & 1 \\ 1 & 0 \\
\end{array} \right)
\end{equation}
and an in\-fini\-tesimal trans\-for\-mation
applied to the fields $\varphi^{a}$ yields
the change
\begin{equation}
\delta \varphi^{a} = \alpha \tau_{b}^{a} \varphi^{b} \ , \
\mid \alpha \mid << 1 \ ,
\end{equation}
The associated SO(1,1) Noether cur\-rent is
\begin{equation}
j^{\nu} = \epsilon_{ab} \varphi^{a} \partial^{\nu} \varphi^{b} \ .
\end{equation}
with $\epsilon_{ab}$ the comp\-letely
anti\-sym\-metric two by two CTP tensor.
Note that $d^{3}x j^{\nu}(x)$ is a dimensionless
quantity.
In the general case, the expectation value of the
Fourier transformed current is given by
\begin{equation}
\label{eq:j}
j^{\nu}(P) = \imath \int \frac{d^{4}\rp}{(2 \pi)^{4}}
\rp^{\nu} \{ G^{21}(P,\rp ) - G^{12}(P,\rp ) \} \ .
\end{equation}
Its zero'th component is related to the
cano\-nical com\-mutator by
\begin{equation}
\label{eq:com}
j^{0}({\vec{x}},t) = \lim_{ {\vec{y}} \rightarrow {\vec{x}} }
\langle \ [ \ {\hat{\varphi}}({\vec{x}},t) \ , \
{\hat{\pi}}({\vec{y}},t) \ ] \ \rangle \ ,
\end{equation}
so that we have
\begin{equation}
j^{0}(P) =  \int \frac{d^3{\vec{\rp}}}{(2 \pi)^4} \delta^{4}(P)
\end{equation}
from canonical quantization.
It implies that $j^{0}(X)$ is an infinite constant.
To see the meaning of this,
consider the expectation
value of the commutator of two fields
\begin{equation}
\langle [ {\hat{\varphi}}(x) , {\hat{\varphi}}(y) ] \rangle
= G^{21}(x,y) - G^{12}(x,y) .
\end{equation}
It represents a pulse arriving at the
point $x = y$ from the backward lightcone,
and subsequently going out into the forward
lightcone \cite{Dir}.
The temporal charge density is the amplitude
corresponding to this,
summed over all possible momenta.
Put differently,
$j^{0}(x)$ is the amplitude density for absorbing
a particle at x and subsequently emiting it.
That it is infinite is due to the fact that it is
obtained from
\begin{equation}
\frac{\partial}{\partial \rx^{0}}\{ G^{21} (X,\rx ) - G^{12}(X,\rx ) \}\ ,
\end{equation}
evaluated at $\rx^{0} = 0$.
It represents the amplitude density for
absorbing a particle and reemmiting it
at a relative distance $\rx$.
Physically,
at $\rx^{0} =0$ the only contribution
can come from ${\vec{\rx}} = 0$,
yielding the delta function.
The infinity is a result of the expansion
of these amplitudes in terms of plane waves.
The use of extended wavepackets on the other
hand would yield finite expressions.
We see that the propagation of particles
is represented by subsequent absorptions
and reemmisions.
In the free field vacuum this amplitude
is conserved.
The temporal charge density is a creation/
annihilation density,
whose free field vacuum value describes
the free propagation of particles.
It is sensible to make a vacuum subtraction
setting the free field temporal charge equal
to zero,
which corresponds to normal ordering.

\par
Analyzing the eigenstates
of the SO(1,1) generator sheds light
on the interpretation of
$j^{0}(P)$ in terms of particles.
The fields
\begin{equation}
{\hat{\varphi}}^{\pm} = \frac{1}{\sqrt{2}}
\{ {\hat{\varphi}}^{1} \pm {\hat{\varphi}}^{2} \}
\end{equation}
are normalised eigen\-vectors of the
SO(1,1) generator.
These are the genera\-tors of the Keldysh
base i.e. the physical
re\-presen\-tation.\cite{Chin}.
We as\-sign the eigen\-values
$q_{\pm} = \pm 1$, or so-called
temporal charge, to these.
With respect to particle/anti-particle
exchange, the states generated by
${\hat{\varphi}}^{+}$ are symmetric and
those generated by ${\hat{\varphi}}^{-}$
are anti-\-sym\-metric.
Since, for neutral bo\-sons, par\-ticles and
anti-\-par\-ticles are in\-distiguishable,
the 1-particle, negative temporal charge states
are redundant.
In the physical limit,
the field carrying negative
temporal charge $< {\hat{\varphi}}^{-}>$
vanishes.
The field-operator ${\hat{\varphi}}^{+}$
creates a symmetric state out of the vacuum,
representing a physical particle.
\par
Let us see if the propagator is
consistent with the demand that
physical particles allways move
into the forward light-cone.
The propagator is given by
${\cal{T}}_{p} \{ {\hat{\varphi}}^{-}
{\hat{\varphi}}^{+} \}$.
By considering the retar\-ded and
advan\-ced 2-point Green functions,
given by
\begin{equation}
\left\{
\begin{array}{lcr}
G_{ret}(x,y) & = & \langle \varphi^{+}(x) \varphi^{-}(y) \rangle \\
G_{adv}(x,y) & = & \langle \varphi^{+}(y) \varphi^{-}(x) \rangle
\end{array}
\right.
\end{equation}
we can identify the propagators for the
particles with + and -
temporal charge respectively.
They represent respectively the
outgoing and the incoming particles.
In terms of these functions we can write
\begin{equation}
j^{\nu}(P) = \int \frac{ d^{4}\rp }{ (2\pi)^{4}} \rp^{\nu}
\{ G_{ret}(P,\rp ) - G_{adv}(P,\rp ) \}
\end{equation}
showing that the temporal current is
directly related to the absorptive
part of the 2-point Green
function \cite{Chin}.
\par
Thus we find that
\begin{equation}
\{ G^{21} (P,\rp ) - G^{12}(P,\rp ) \}
\end{equation}
is the creation/annihilation density,
in $(P,\rp )$-space,
corresponding to momentum-loss $P$
and the momentum-flow $\rp$.
Since we have lost all coordinate space
information,
we cannot say in what region of coordinate space
the momentum flows and where it is lost.
However such questions can be addressed by
constructing appropriate wave-packets.
The commutation relation Eq.(\ref{eq:com})
expresses that if one integrates out the
energy-flow,
then $P=0$.
Put differently,
the momentum is lost to
other particles with different energies
but it is not destroyed.
The temporal current is
the total flow of the creation/\-
annihilation density
in terms of either a location,
$j^{\nu}(X)$,
or the energy-momentum
dissipation $j^{\nu}(P)$.
In detailed balance,
the vacuum subtracted flow will be zero.
The total temporal charge
\begin{equation}
Q(t) = \int d^{3}x \ \epsilon_{ab} \varphi^{a} \partial^{0} \varphi^{b} \ ,
\end{equation}
at time t is the total creation/
\-annihilation amplitude.
Its Fourier transform is
\begin{equation}
Q(e) = \int dt e^{\imath e t} Q(t)
\end{equation}
the amplitude associated with
the energy-loss e.
The divergence of the temporal current
is thus the amplitude for the
energy-momentum loss $P$ due to
all interactions.
\par
The divergence of the temporal current
is related to the integrated
kinetic equations for $G^{21}$ and
$G^{12}$ since
\begin{equation}
P_{\nu}j^{\nu}(P) = \int \frac{d^4 \rp}{(2 \pi)^4}
\ \rp_{\nu}P^{\nu} \{ G^{21} (P,\rp ) -  G^{12} (P,\rp ) \} \ .
\end{equation}
As expected,
and as can be seen from Eq.(\ref{eq:ke}),
the global SO(1,1) symmetry is bro\-ken by the
$\lambda\varphi^{4}$ self-\-inter\-action.
There is, however, an\-other source of SO(1,1)
sym\-metry brea\-king.
Consider the free-\-field,
vacuum,
2-\-point fun\-ctions.
If SO(1,1) sym\-metry were realized
then $G^{ab}$ would be expressible as
\begin{equation}
G_{sym}^{ab}(\rp ) = g_{1}c^{ab} + g_{2}\epsilon^{ab} \ .
\end{equation}
with $g_{1}$ and $g_{2}$ functions of the momentum $\rp$.
The actual Green function can be decomposed
however to display the form
\begin{equation}
G_{vac}^{ab} = (\rp^{2} - m^{2} -\imath \alpha)^{-1} c^{ab}
- 2\pi \imath \theta(\rp^{0} ) \delta ( \rp^{2} - m^{2} )  \epsilon^{ab}
- 2\pi \imath \delta ( \rp^{2} - m^{2} ) \left(
\begin{array}{rr} 0 & 0 \\ 1 & 0 \end{array} \right)
\end{equation}
where one sees that the third term explicitly
breaks SO(1,1) sym\-metry.
A pro\-per de\-finition of the
non-\-equi\-librium theory,
i.e. fixing the initial con\-ditions,
also re\-quires the sym\-metry to be bro\-ken.

\subsection{Initial Conditions Breaking SO(1,1)}

\indent
For the sake of sim\-plicity,
we examine the sym\-metry brea\-king
by a 2-point kernel $K_{ab}(x,y)$.
This covers both
equi\-librium and non-\-equilibrium
initial conditions,
that describe an in\-homogeneous sys\-tem
of free bo\-sons.
If $\lambda = 0$,
then the di\-vergence
of the tem\-poral cur\-rent is
\begin{equation}
\label{eq:div}
\partial_{\nu} j^{\nu}(x) = \int d^{4}y \ \ K_{ab}(x,y) \varphi^{a}(y)
\tau_{c}^{b} \varphi^{c}(x) \ .
\end{equation}
If this is integrated over $x$,
after vacuum subtraction,
we obtain the the total particle
production amplitude,
\begin{equation}
N_{tot} = K_{ab} \circ \{ {\cal{G}}^{ab} - {\cal{G}}_{vac}^{ab} \} \ .
\end{equation}
These particles are created to
form the initial state.
Taking the most general form for
the sym\-metry-brea\-king ker\-nel
to be
\begin{equation}
K_{ab}(x,y) = k_{1}(x,y) \delta_{ab} + k_{2}{\tilde{\tau}}_{ab}
\end{equation}
with
\begin{equation}
{\tilde{\tau}}_{ab} = \tau_{b}^{a} ,
\end{equation}
and using
\begin{equation}
G^{21} + G^{12} = G^{11} + G^{22} = G_{corr} ,
\end{equation}
we see the di\-vergence of the tem\-poral
cur\-rent is di\-rectly related to the
2-point cor\-relation function
\begin{equation}
\label{eq:div2}
\partial_{\nu}j^{\nu}(x) = \int d^{4}y \{ k_{1}(x,y)
+ k_{2}(x,y) \} G_{corr}(x,y) \} \ .
\end{equation}
In terms of temporal charge fields
we have
\begin{equation}
G_{corr}(x,y) = \langle \varphi^{+}(x) \varphi^{+}(y) \rangle \ ,
\end{equation}
re\-presen\-ting the cor\-relation among
the par\-ticle pro\-duction at $x$
and at $y$.
So far we have discussed the expectation values of the
SO(1,1) current operator.
However in the context of Quantum
Field Theory symmetries,
even explicitly broken symmetries,
generate a whole hierarchy
of identities among Green
functions.
\par
By ap\-plying the usual tech\-niques to
the func\-tional integral,
as\-suming the ab\-sence of ano\-malous
brea\-king of the SO(1,1) sym\-metry,
we can derive Ward-\-Takahashi identities
for the broken sym\-metry.
In the case of initial con\-ditions
symmetry-\-breaking (ICSB) dis\-cussed
above we find
\begin{eqnarray}
\label{eq:ward1}
\frac{\partial}{\partial z^{\nu}} \ \langle
j^{\nu}(z) \prod_{j=1}^{n} \varphi^{a_{j}}(x_{j}) \rangle &
= &
\sum_{j=1}^{n} \langle \delta^{4}(z - x_{j} ) \tau_{c}^{a_{j}} \varphi^{c}(z)
\prod_{i \neq j}^{n} \varphi^{a_{i}}(x_{i}) \rangle \nonumber \\
& + &
\int d^4y \ K_{ab}(z,y) \langle \varphi^{a}(z) \varphi^{b}(y)
\prod_{j=1}^{n} \varphi^{a_{j}}(x_{j}) \rangle \ , \nonumber \\
\end{eqnarray}
where the braces indicate func\-tional integral averages.
For $n=0$ we re\-cover Eq.(\ref{eq:div}).
For $n>0$ we ex\-pect the first term
to become small as $z \rightarrow \infty$
and the aapplication of Gauss theorem
to the l.h.s. of Eq.(\ref{eq:ward1})
leads to the relation
\begin{equation}
\label{eq:intwrd1}
\sum_{j=1}^{n} \tau_{c}^{a_{j}} \langle \varphi^{c}(x_{j}
\prod_{i\neq j}^{n} \varphi^{a_{i}}(x_{i}) \rangle
=
- K_{ab} \circ \langle \varphi^{a} \varphi^{b}
\prod_{j=1}^{n} \varphi^{a_{j}}(x_{j}) \rangle \ .
\end{equation}
For n=1,2 we have explicitly
\begin{equation}
\mfi{a}(x) = -K_{bc}(y,z) \circ {\cal{G}}^{bca}(y,z,x) \ ,
\end{equation}
and
\begin{equation}
\label{eq:cor}
G_{cor}(x,y) =  - K_{bc}(z,z') \circ {\cal{G}}^{bc11}(z,z',x,y) ,
\end{equation}
re\-presenting the con\-tribution of the
initial con\-ditions to par\-ticle
dis\-tribu\-tions at later times.
\par
We have seen how initial con\-ditions break
SO(1,1) sym\-metry
in order to pre\-pare the initial state,
and we have cal\-culated the
effect this has on the CTP Green
func\-tions.
The self-\-inter\-action also breaks
the sym\-metry and we will deal with that
next.

\subsection{Interactions Breaking SO(1,1)}

\indent
The divergence in the temporal
current due to inter\-actions is
\begin{equation}
\label{eq:div2}
\partial_{\nu} j^{\nu}(x) = -\frac{\lambda}{3!}h_{abcd} \tau_{e}^{d}
\varphi^{a}(x)\varphi^{b}(x)\varphi^{c}(x)\varphi^{e}(x)  \ .
\end{equation}
This ex\-plicit sym\-metry brea\-king (ESB)
is local and there\-fore per\-sistent,
un\-like the case dis\-cussed above.
A simple example of the vio\-lation
of particle num\-ber by inter\-actions
is the de\-cay of one boson into three bosons.
Note that Eq.(\ref{eq:div2}) is just
the $d^{4}\rp $ integra\-ted kinetic equa\-tion
for $G^{21}-G^{12}$.
The integrated Vlasov- and collision terms can all be
recovered from Eq.(\ref{eq:div2}),
and interpreted along the lines of (off-shell)
particle creation and annihilation in the
vacuum of the interacting theory.
\par
For ESB we also give a Ward-\-Takahashi identity,
assuming no ICSB one finds
\begin{eqnarray}
\label{eq:ward2}
\frac{\partial}{\partial z^{\nu}} \ \langle
j^{\nu}(z) \prod_{j=1}^{n} \varphi^{a_{j}}(x_{j}) \rangle
& = &
\sum_{j=1}^{n} \langle \delta^{4}(z - x_{j} ) \tau_{c}^{a_{j}} \varphi^{c}(z)
\prod_{i \neq j}^{n} \varphi^{a_{i}}(x_{i}) \rangle \nonumber \\
& - &
\frac{\lambda}{3!}h_{abcd} \tau_{e}^{d} \langle
\varphi^{a}(z)\varphi^{b}(z)\varphi^{c}(z)\varphi^{e}(z)
\prod_{j=1}^{n} \varphi^{a_{j}}(x_{j}) \rangle \ .
\end{eqnarray}
Once again for $n>0$,
we ex\-pect an
in\-tegrated identity to hold.
We find
\begin{equation}
\sum_{j=1}^{n} \tau_{c}^{a_{j}} {\cal{G}}^{a_{1} ...a_{j-1} c a_{j+1}... a_{n}}
=
\frac{\lambda}{3!}h_{abcd} \circ \tau_{e}^{d} {\cal{G}}^{abcea_{1} ... a_{n}} \
{}.
\end{equation}
Again we give to explicit examples,
for $n=1$
\begin{equation}
\mfi{a} = \frac{\lambda}{3!} h_{bcde} \circ
\tau^{a}_{f} \tau^{e}_{g} {\cal{G}}^{bcdgf} \ ,
\end{equation}
and for $n=2$, we find an equation for
the correlation function
\begin{equation}
G_{corr}(x,y) = \frac{\lambda}{3!}h_{abcd} \tau^{d}_{e}
\circ {\cal{G}}^{abce11} \ .
\end{equation}
In any rea\-listic case,
both mecha\-nisms of SO(1,1) brea\-king con\-tribute.
The bro\-ken SO(1,1) sym\-metry ge\-nerates a
hier\-archy of equa\-tions for the
Green func\-tions.
The generaliza\-tion of Eq.(\ref{eq:ward1})
and Eq.(\ref{eq:ward2})
to higher-\-order ker\-nels $K_{abc...}$
and other poly\-nomial inter\-actions
is straight\-forward.

\newpage

\section{Energy-Momentum Conservation}

\indent
In this sec\-tion,
we show that the di\-vergence of the
ca\-nonical energy-\-momentum
ten\-sor vanishes due to the ki\-netic and
con\-straint equa\-tions.
It will turn out that the true dif\-feren\-ce
bet\-ween Vlasov- and col\-lision-\-terms
lies in the (non-)\-loca\-lity of the inter\-actions
that they re\-present.
\par
The energy-\-momentum tensor in the $\lambda \varphi^{4}$
model is given by
\begin{eqnarray}
T^{\mu \nu} & = & \partial^{\mu} \varphi \partial^{\nu} \varphi
- \eta^{\mu \nu} \frac{1}{2} \{ \partial_{\alpha} \varphi \partial^{\alpha}
\varphi - m^{2}\varphi^{2} - 2 \frac{\lambda}{4!}\varphi^{4} \} \nonumber \\
  &  =  & T_{0}^{\mu \nu} + \frac{\lambda}{4!}T_{int}^{\mu \nu} \ ,
\end{eqnarray}
where $T_{0}^{\mu \nu}$ refers to the free field
energy-\-momentum ten\-sor and $T_{int}^{\mu \nu}$ to the
inter\-acting part.
In terms of 2-point func\-tions this can be written as
\begin{equation}
\label{eq:Gem}
\langle T_{0}^{\mu \nu} \rangle = \frac{1}{2} \int \frac{d^{4}\rp}{(2\pi )^{4}}
\{ -\frac{1}{4}P^{\mu}P^{\nu} + \rp^{\mu}\rp^{\nu} - \frac{1}{2} \eta^{\mu \nu}
(\rp^{2} - \frac{1}{4}P^{2} - m^{2} ) \}
G_{corr}(P,\rp ) \ ,
\end{equation}
where $G_{corr}$ has been defined in Eq.(3.23).
In deriving this expression,
linear terms in $\rp^{\nu}$ have been dropped,
because they do not contribute to the integral
since $G_{corr}(P,\rp )$ is symmetric in $\rp^{\nu}$.
The interaction contribution to the
energy momentum tensor is connected
with the coincidence limit of the
four point Green functions.
One possible choice is
\begin{equation}
\langle T_{int}^{\mu \nu} \rangle = \frac{1}{2} \eta^{\mu \nu}
{\cal{G}}^{1122}(x,x,x,x) + {\cal{G}}^{1211}(x,x,x,x) \ .
\end{equation}
It is important to recognize that
the interaction part also contains
the contribution of the {\it independent \/}
variable $\alpha_{abcd}^{4}$.
Energy-\-momentum conservation requires
\begin{equation}
\label{eq:emc}
P_{\mu} \langle T_{0}^{\mu \nu} \rangle =
\frac{\lambda}{4!}P_{\mu}\langle T_{int}^{\mu \nu} \rangle \ .
\end{equation}
We will investigate this relation for three
cases.
First we will consider the free field situation,
followed by a verification of Eq.(\ref{eq:emc})
taking into account the Vlasov interaction and,
finally, treating the full theory.
Our procedure will be to use the kinetic and
constraint equations to calculate the l.h.s. of
Eq.(\ref{eq:emc}) from Eq.(\ref{eq:Gem}) and
conclude the equality with the r.h.s. of
Eq.(\ref{eq:emc}).
\par
For the free field case, $\lambda = 0$,
the kinetic and constraint equations read
\begin{equation}
\left\{
\begin{array}{lc}
\rp . P G^{ab}(P,\rp  ) = 0 &  \\
\{ {\rp}^{2} + \frac{1}{4} P^{2}- m^{2} \} G^{ab}(P,\rp )
= & c^{ab} (2 \pi)^{4} \delta^{4}(P) \ .
\end{array}
\right.
\end{equation}
Applying $P_{\mu}$ to Eq.(\ref{eq:Gem}),
it is straightfor\-ward to see
that the l.h.s. of Eq.(\ref{eq:emc})
vanishes.
In the Vlasov ap\-proximation,
we neg\-lect
the col\-lision terms, i.e. we treat
the 4-\-point Green func\-tion as a
dis\-con\-nected pro\-duct of 2-\-point
func\-tions.
So the inter\-action term is
\begin{equation}
\label{eq:bc}
\langle T_{int}^{\mu \nu} \rangle = \frac{3}{2} \eta^{\mu \nu}
\{ G^{11}(x,x)G^{12}(x,x) + G^{22}(x,x)G^{21}(x,x) \} \ .
\end{equation}
The factor 3 counts the possible ways of
writing a 4-point Green function as a disconnected
product of 2-point Green functions.
The ki\-netic and con\-straint equa\-tions
are now given as
\begin{equation}
\left\{
\begin{array}{lc}
\imath \rp . P G^{ab}(P,\rp  )
- \frac{\lambda}{2} \int \frac{d^{4}P' }{(2 \pi )^{4}}
\frac{d^{4} \rp '}{(2 \pi )^{4}} G^{aa}(P',\rp ')
\sinh \{ \frac{1}{2} P'. \partial_{\rp } \} G^{ab}(P-P',\rp )
= 0 & \\
\imath \{ {\rp}^{2} + \frac{1}{4} P^{2}- m^{2} \} G^{ab}(P,\rp )
- \frac{\lambda}{2} \int \frac{d^{4}P'}{(2 \pi )^{4}}
\frac{d^{4}\rp '}{(2 \pi )^{4}}
G^{aa}(P',\rp ')\cosh \{
\frac{1}{2} P' . \partial_{\rp } \} G^{ab}(P-P',\rp ) & \\
 =  \imath c^{ab} (2 \pi)^{4} \delta^{4}(P) & \ .
\end{array}
\right.
\end{equation}
Applying $P_{\mu}$ to Eq.(\ref{eq:Gem}) and using
the Vlasov equations yields
\begin{eqnarray}
P_{\mu}\langle T_{0}^{\mu \nu} \rangle (P) & =
& \frac{1}{2} \int \frac{d^{4}\rp}{(2 \pi )^{4}}
\int \frac{d^{4}P' }{(2 \pi )^{4}} \frac{d^{4} \rp '}{(2 \pi )^{4}} \nonumber
\\
& \rp^{\nu} \{ & G^{22}(P',\rp ') \sinh \{ \frac{1}{2} P'. \partial_{\rp } \}
G^{21}(P-P',\rp )  \nonumber \\
& + & G^{11}(P',\rp ') \sinh \{ \frac{1}{2} P'. \partial_{\rp } \}
G^{12}(P-P',\rp ) \} \nonumber \\
+ & \frac{P^{\nu}}{2} \{ & G^{22}(P',\rp ')\cosh \{ \frac{1}{2} P' .
\partial_{\rp } \}
G^{21}(P-P',\rp ) \nonumber \\
& + & G^{11}(P',\rp ')\cosh \{ \frac{1}{2} P' . \partial_{\rp } \}
G^{12}(P-P',\rp ) \} \ . \nonumber \\
\end{eqnarray}
The $d^{4}\rp$-integration in the terms
containing the sinh and cosh functions
can be performed.
For the sinh terms this yields
\begin{equation}
\int \frac{d^{4}\rp}{(2 \pi )^{4}}
\rp^{\nu} G^{aa}(P',\rp ')
\ \sinh \{ \frac{1}{2} P'. \partial_{\rp } \}
G^{ab}(P-P',\rp ) = - \frac{P^{\nu '}}{2} G^{aa}(P',\rp ') \int
\frac{d^{4}\rp}{(2 \pi )^{4}} G^{ab}
(P-P', \rp) ,
\end{equation}
to be integrated over $P'$ and $\rp '$.
For the cosh terms one can use
\begin{equation}
\int \frac{d^{4}\rp}{(2 \pi )^{4}}
\frac{P^{\nu}}{2} G^{aa}(P',\rp ')
\ \cosh\{ \frac{1}{2} P'. \partial_{\rp } \}
G^{ab}(P-P',\rp ) = - \frac{P^{\nu '}}{2} G^{aa}(P',\rp ') \int
\frac{d^{4}\rp}{(2 \pi )^{4}} G^{ab}
(P-P', \rp) .
\end{equation}
These terms generate the r.h.s. of Eq.(\ref{eq:emc}).
Note that, since the Vlasov term has
gra\-dient opera\-tors bet\-ween a
$\rp$-\-indepen\-dent factor and
$G^{ab}(P,\rp )$, the gra\-dients
play no role in energy-\-momen\-tum
con\-ser\-vation.
\par
In the ful\-ly inter\-acting case, the
con\-fir\-mation of Eq.(\ref{eq:emc}) be\-comes
difficult as long as we stay in
$(P,\rp )$-space.
Sub\-stitu\-tion of Eq.(\ref{eq:ke}) and Eq.(\ref{eq:ce})
into the l.h.s. of Eq.(\ref{eq:emc}) now yields
\begin{eqnarray}
\label{eq:ab}
P_{\mu}\langle T_{0}^{\mu \nu}(P) \rangle &
= \frac{1}{2} \frac{\lambda}{4!}
  \int \frac{d^{4}\rp }{(2\pi )^{4}}
  \frac{d^{4}P'}{(2\pi )^{4}}
  \rp^{\nu} \{
   \Sigma_{\ c}^{2}(P-P',\rp) \Delta^{-} G^{c1}(P',\rp ) +
   \Sigma_{\ c}^{1}(P-P',\rp) \Delta^{-} G^{c2}(P',\rp ) & \nonumber \\
& -  G_{\ c}^{2}(P-P',\rp) \Delta^{+} \Sigma^{c1}(P',\rp ) -
    G_{\ c}^{1}(P-P',\rp) \Delta^{+} \Sigma^{c2}(P',\rp )
    \} & \nonumber \\
&  \frac{P^{\nu}}{2} \{
   \Sigma_{\ c}^{2}(P-P',\rp) \Delta^{-} G^{c1}(P',\rp ) +
   \Sigma_{\ c}^{1}(P-P',\rp) \Delta^{-} G^{c2}(P',\rp ) & \nonumber \\
& +  G_{\ c}^{2}(P-P',\rp) \Delta^{+} \Sigma^{c1}(P',\rp ) +
     G_{\ c}^{1}(P-P',\rp) \Delta^{+} \Sigma^{c2}(P',\rp )
\} & ,
\end{eqnarray}
where we have allready used the fact that the free
field and Vlasov terms either vanish or are taken care
of by corresponding terms in the interaction part.
The $d^{4}\rp$-integration in Eq.(\ref{eq:ab}) is not
trivial, as it was for the Vlasov term.
All the gradients must be included.
We do this by transforming Eq.(\ref{eq:ab})
back to coordinate-space.
There we have
\begin{eqnarray}
\lim_{y \rightarrow x}
& \frac{1}{2}\frac{\lambda}{4!}\int d^{4}z
& \{\frac{\partial}{\partial x^{\nu}}
[ \Sigma_{\ c}^{2}(x,z) G^{c1}(z,y) + \Sigma_{\ c}^{1}(x,z) G^{c1}(z,y) ]
\nonumber \\
& +
& \{\frac{\partial}{\partial y^{\nu}}
[ G_{\ c}^{2}(y,z) \Sigma^{c1}(z,x) + G_{\ c}^{1}(y,z) \Sigma^{c1}(z,x) ] \} \
,
\end{eqnarray}
which we can identify as the point-\-split
implementation of the product rule by
using the symmetry of the Green functions
under the coherent exchange of CTP indices
and the corresponding space-time arguments.
Consequently we have
the divergence of the free
field energy-momentum tensor
expressed in terms of the
connected 4-point Green function
\begin{equation}
\partial_{\mu} \langle T_{0}^{\mu \nu} \rangle = \frac{\lambda}{4!2}
\partial^{\nu}
\{ G^{2221}(x,x,x,x) + G^{1112}(x,x,x,x) \} \ .
\end{equation}
Using Eq.(\ref{eq:bc}) we see that
Eq.(\ref{eq:emc}) is satisfied,
so that in general energy and momentum
are conserved.
Contrary to energy-momentum conservation
for the Vlasov terms we required all
the gradients.
The physical reason for this
lies in the nature of the interactions
that these terms represent.
The Vlasov-\-term results from
{\it local \/ } inter\-actions with
(off-shell) par\-ticles from the back\-ground.
The interaction vertex is the
point-like $\lambda \varphi^{4}$ interaction.
The col\-lision terms con\-tain
{\it non-local \/}inter\-actions,
like, for example,
the two-boson exchange interaction.
The gradients probe the locality of
these interactions.
Neg\-lecting gra\-dients thus vio\-lates
energy-\-momen\-tum con\-servation,
un\-less specific non-\-local inter\-actions are ad\-ded to the
Lagran\-gian density.
\par
Energy-\-momentum con\-servation is
guaranteed by the ki\-netic- and
con\-straint equa\-tions.
This points to the fact that
any in\-homo\-geneities oc\-curring in such a
field theoretic model come from a spon\-taneous
brea\-king of trans\-lation in\-variance.
A gradient-expansion violates locality
if used in the collision term.
A Vlasov-\-ap\-proximation is restricted to
the local response of the system
to inhomogeneities.
The temporal current non-conservation
is a differential equation for
the difference $G^{21} - G^{12}$
in the coincidence limit and
energy-momentum conservation is a
differential equation for the sum
$G^{21} + G^{12}$ in the coincidence limit.
The two equations,
given apropriate boundary conditions,
consequently determine the
coincidence limit of the
2-point Green functions.
Put differently,
energy-momentum conservation and
temporal current non-conservation
determine the U.V.behaviour of
the dressed propagators.
In the next section we analyse
a free field case in the presence of
inhomogeneities.

\section{Free field}

\par
\indent
In the previous sections we have seen
that energy-momentum conservation and
temporal current non-conservation
equations determine the coincidence
limits of $G^{21}(x,y)$ and $G^{12}(x,y)$.
Inhomogeneities have played a major role
in establishing the conservation
since they reflect the energy-momentum
exchange between the particles.
Inhomogeneities on the other hand
make their appearance in the gradient
expansions of the interaction terms
in the kinetic and constraint equations,
but also they appear quadratically in
the free field equations.
In this section we will concentrate
on the free field equations
\begin{equation}
\label{eq:fke}
\left\{
\begin{array}{lc}
\rp . P G^{ab}(P,\rp  ) = 0 & \\
\{ {\rp}^{2} + \frac{1}{4} P^{2}- m^{2} \} G^{ab}(P,\rp )  = & c^{ab} (2
\pi)^{4} \delta^{4}(P)
\end{array}
\right.
\end{equation}
To find the solution to Eq.(\ref{eq:fke})
for arbitrary initial data,
we exploit the linearity of
these equations.
It is used to discuss the effect
of inhomogeneities and
off-shellness.
Following this we give an explicit example
of an inhomogeneous non-equilibrium system.

\subsection{General solution}

\indent
By inspection of Eq.(\ref{eq:fke}) one
sees that the decomposition of $G^{ab}$
into a vacuum contribution,
$G_{vac}^{ab}$,
and a medium term,
${\tilde{G}}^{ab}$,
\begin{equation}
\label{eq:dd}
G^{ab}(P,\rp ) = G_{vac}^{ab}(P,\rp ) + {\tilde{G}}^{ab}(P,\rp ) \ ,
\end{equation}
solves the free field equations,
where $G_{vac}^{ab}$ is
\begin{equation}
G_{vac}^{ab}(P,\rp ) = (2 \pi )^{4} c^{ab}
\frac{\delta^{4}(P^{\nu}) }{ \rp^{2} + \frac{1}{4}P^{2} - m^{2} }
\end{equation}
and ${\tilde{G}}^{ab}(P,\rp )$ satisfies
the set of homogeneous equations
\begin{equation}
\label{eq:hom}
\left\{
\begin{array}{lc}
\rp . P {\tilde{G}}^{ab}(P,\rp  ) = 0 & \\
\{ {\rp}^{2} + \frac{1}{4} P^{2}- m^{2} \} {\tilde{G}}^{ab}(P,\rp )  = & 0 \ .
\end{array}
\right.
\end{equation}
Since $G_{vac}^{ab}$ describes
the propagation of free particles,
their $P$-dependence is a simple $\delta^{4}(P)$
as expected from the absence of interactions.
All the momentum-loss information is included
in ${\tilde{G}}^{ab}$.
It is the momentum-loss suffered by the
particles during the preparation of
the initial condition.
Particle propagation at later times
will not couple to these initial
inhomogeneities and thus not
dissipate them.
\par
Let us consider that $G^{ab}(X,\rx )$
is given by an initial value
at $X^{0} = 0 \ and \ \rx^{0} = 0$ and denote
it as $G_{0}^{ab}(X^{i},\rx^{j})$.
Since we work in momentum space,
it is useful to introduce the Fourier
transform of the initial conditions
$G_{0}^{ab}(P^{i},\rp^{j})$
\begin{equation}
{\tilde{G}}_{0}^{ab}(P^{i},\rp^{j}) = \int d^{3}X \int d^{3}\rx
e^{- \imath P_{i}X^{i} - \imath \rp_{j}\rx^{j} }
{\tilde{G}}_{0}^{ab}(X^{i},\rx^{j}) \ .
\end{equation}
This Fourier transform can also be decomposed
in the fashion of Eq.(\ref{eq:dd}) as
$G_{0}^{ab} = G_{0 \ vac}^{ab}
+ {\tilde{G}}_{0}^{ab}$.
The linearity of the equations implies the solution
for ${\tilde{G}}^{ab}$ has the form
\begin{equation}
{\tilde{G}}^{ab}(P,\rp ) = {\tilde{G}}_{0}^{ab}({\vec{P}},{\vec{\rp }})
\delta (\rp_{\nu}P^{\nu}) \delta ( \rp^{2} + \frac{1}{4} P^{2} - m^{2} ) \ .
\end{equation}
Due to the symmetry,
$G^{21}(P,\rp ) = G^{12}(P,-\rp )$,
of the off-diagonal 2-point functions
and the relation
\begin{equation}
G^{11}(X,\rx ) = \theta (\rx^{0}) G^{21}(X,\rx ) + \theta (-\rx^{0})
G^{12}(X,\rx ) \ ,
\end{equation}
specifying the diagonal elements in terms of
the off-diagonal elements,
the four functions ${\tilde{G}}^{ab}$
contain only one unknown function $g(P,\rp )$,
satisfying Eq.(\ref{eq:hom}).
In thermal equilibrium,
it is well-knowm that
$g(P,\rp)$ is given as
\begin{equation}
g(P,\rp ) = g_{\small BE}(p^{0}) \delta (\rp^{2} - m^{2} ) \delta^{4}(P) \ ,
\end{equation}
where $g_{\small BE}(p^{0} )$ is the
Bose-Einstein distribution function.
\par
Consider the Wigner representation
of the general solution
\begin{equation}
\label{eq:genwig}
g(X,\rp ) = \int \frac{d^{4}P}{(2\pi)^{4}} e^{\imath P_{\mu}X^{\mu}}
\delta (\rp_{\mu}P^{\mu} ) \delta ( \rp^{2} + \frac{1}{4} P^{2} - m^{2} )
g(P,\rp)
\end{equation}
Since we deal with massive particles
$\rp^{0} \neq 0$
and two of the four $P$ integrations,
say over $P^{0}$ and the component
parallel to $\rp^{j}$
\begin{equation}
P_{\parallel}^{j} = \frac{ P_{i}\rp^{i} }{ \rp_{i}\rp^{i} }\rp^{j}
\ ,
\end{equation}
can be performed using
\begin{equation}
\label{eq:id2}
\left\{
\begin{array}{lcr}
\delta(\rp_{\mu}P^{\mu}) & = & \frac{1}{\mid \rp^{0} \mid} \delta (P_{0} -
 \frac{ \rp P_{\parallel} }{ \rp^{0} }) \\
\delta ( \rp^{2} + \frac{1}{4} P^{2} - m^{2} ) & = &
 \frac{1}{\mid \gamma \alpha \mid} \{
 \delta (P_{\parallel} - 2 \gamma \alpha) +
 \delta (P_{\parallel} + 2 \gamma \alpha) \}
\end{array}
\right.
\end{equation}
where we have defined the measure of
off-shellness $\alpha$ by
\begin{equation}
\label{eq:al}
\alpha = \sqrt{ \rp^{2} - m^{2} - \frac{1}{4} P_{\perp}^{2} }
\end{equation}
and
\begin{equation}
\label{eq:lb}
\gamma = \frac{1}{ \sqrt{1 - \frac{ \rp_{i}\rp^{i} }{\rp_{0}^{2} } }} \ ,
\end{equation}
and used the first identity in Eq.(\ref{eq:id2})
to obtain the second.
In these coordinates $\rp$ has only two
non-vanishing elements; $\rp^{0}$ and $\rp $.
Thus $\gamma$ is a Lorentz
boost factor and $\alpha$
is {\it not \/} a Lorentz-invariant
quantity, but rather a frequency.
After performing the integrals in
$P^{0}$ and $P_{\parallel}$,
the integrand will have the general form
\begin{equation}
\label{eq:genres}
g(X,\rp,\rp^{0}) = \int \frac{ d^{2}P_{\perp} }{ (2\pi)^{2} }
W(\rp,\rp^{0},\gamma,\alpha,X_{\perp}) e^{\imath 2 \gamma \alpha [
X_{\parallel} -
\frac{\rp}{\rp^{0} } X_{0} ] } +
W(\rp,\rp^{0},\gamma,\alpha,X_{\perp}) e^{-\imath 2 \gamma \alpha [
X_{\parallel} -
\frac{\rp}{p^{0} ] } X_{0} }  \ .
\end{equation}
Eq.(\ref{eq:genres}) is a superposition
of waves with frequencies $\gamma \alpha$
moving along the $X_{\parallel}$-axis.
The Lorentz boost factor relates these waves
to spatial oscillations in the
rest-frame defined by ${\vec{p}}=0$.
\par
Although there is no interaction among
the particles,
there are collective off-shell effects
due to the preparation of the initial condition.
In a free field theory,
particles with a pure momentum are
described by plane waves.
Eq.(\ref{eq:genres}) tells us that a particle
cannot be in a pure momentum state if there
are transversal inhomogeneities (T.I.'s).
Clearly they would disturb any plane wave
into a more complex wave.
{}From the reality condition on $\alpha$,
we see that T.I.'s must vanish for $\rp$
to be on-shell.
The constraint $\rp_{\mu}P^{\mu} = 0$ implies that
the rest-mass of the particles is un\-affected
by inhomogeneities.
The energy-loss is solely due to
longitudinal inhomogeneities (L.I.'s),
in contrast to the T.I.'s, responsible
for removing the particles off the mass shell.
By going to the rest frame $\rp = 0$ of the
particles moving along $X_{\parallel}$,
one sees that $\rp^{0} > m$ for non-vanishing T.I.'s.
This implies that the particles will not
reach asymptotic on-shell states unless
they were not deflected in the first place.
One of the two remaining integrations over
the components of $P_{\perp}$ can be replaced
by an integration over the off-shellness.
Integrating out the $\alpha$
represents a summing over quantum
fluctuations dressing the particles.

\subsection{Free expansion of a Bose-Einstein gas}

\indent
In this subsection, we give an
example of a system as discussed above.
The free expansion of a Bose-Einstein
gas will be treated.
Consider $(3+1)$ dimensional Minkowski
space divided into two half-spaces,
$M_{+}$ and $M_{-}$, by a
Dirichlet-wall at $x^{3}=0$.
The gas in $M_{-}$ is assumed to be in
thermal equilibrium at a temperature
T, while in $M_{+}$ a vacuum persists.
The total-space propagator $G(x,y)$
is defined as a sum of two half-space
propagators
\begin{equation}
\label{eq:def}
G(x,y) = \theta (x^{3}) \theta (y^{3}) G_{+}(x,y)
+  \theta (- x^{3}) \theta (- y^{3}) G_{-}(x,y)
\end{equation}
where $G_{+}$ is a vacuum propagator
and $G_{-}$ is a thermal propagator.
These can be computed from total-space
vacuum and thermal propagators by the
method of images.
This procedure can be justified by
an eigenfunction expansion in terms
of solutions of the half-space
Klein-Gordon equation.
For the half-space vacuum,
one can express the associated Green function
in terms of the total-space vacuum propagator
$G_{vac}(x,y)$
\begin{equation}
G_{+}(x,y) = G_{vac}(x;y) - G_{vac}(x^{i},-x^{3};y) \ .
\end{equation}
By using the fact that the reflection
$(x^{3},y^{3}) \rightarrow (-x^{3},y^{3})$,
is equivalent to $(X^{3},\rx^{3}) \rightarrow
(-\frac{1}{2}\rx^{3}, - 2 X^{3})$,
in the centre of mass (c.m.)
and relative coordinates,
we can write down the relation
in the Fourier representation as
\begin{equation}
G_{+}(P;\rp) = G_{vac}(P;\rp) -
G_{vac}(P^{i},-2\rp^{3};\rp^{i},-\frac{1}{2}P^{3}) \ .
\end{equation}
In the thermal half-space on the other
hand we have,
using the thermal total-space propagator
$G_{T}(P,\rp )$,
\begin{equation}
G_{-}(P;\rp) = G_{T}(P;\rp) -
G_{T}(P^{i},-2\rp^{3};\rp^{i},-\frac{1}{2}P^{3}) \ .
\end{equation}.
These propagators describe a stationary
state which at $X^{0}=\rx^{0}=0$ coincides
with our initial condition.
In momentum space the initial conditions
in each of the half-spaces is now
represented by the apropriate propagator
integrated over $P^{0}$ and $\rp^{0}$.
Thus we compute the initial condition
for each half-space
\begin{equation}
G_{\pm \ 0}(P^{i},\rp^{j})
= \int \frac{dP^{0}}{2 \pi}\frac{d\rp^{0}}{2 \pi}
G_{\pm}(P,\rp) \ .
\end{equation}
The total space initial condition
is calculated from the functions $G_{\pm \ 0}$
according to Eq.(\ref{eq:def}).
The product with the step functions in
coordinate space becomes a convolution in
momentum space.
The Fourier transform of the product of step\-
functions is found by noting that
\begin{equation}
\theta (\pm x^{3})\theta (\pm y^{3}) =
\theta (\pm X^{3}) \{ \theta (\pm 2X^{3} - \rx^{3})
- \theta(\mp 2 X^{3} - \rx^{3} ) \} \ ,
\end{equation}
and Fourier transforming this yields
\begin{equation}
{\tilde{\theta}}_{\pm}(P,\rp ) \equiv \frac{
\pm 4\rp^{3}}{ [\rp^{3} + \imath {\hat{\epsilon}} ] [
(P^{3})^{2} - 4 (\rp^{3})^{2} \pm \imath 2 P^{3}\epsilon  - \epsilon^{2} ] } \
,
\end{equation}
where the limit ${\hat{\epsilon}},\epsilon \rightarrow 0$
is understood.
Thus the required full-space initial
conditions are given by
\begin{eqnarray}
\label{eq:con}
G_{0}(P^{i},\rp^{j}) & = &
\int \frac{dP_{3}'}{2\pi} \frac{d\rp_{3}'}{2\pi}
{\tilde{\theta}}_{+}(P',\rp') G_{+ \ 0}(P_{i} - P_{i}',\rp_{j} - \rp_{j})
\nonumber \\
& + & \int \frac{dP_{3}'}{2\pi} \frac{d\rp_{3}'}{2\pi}
{\tilde{\theta}}_{-}(P',\rp') G_{- \ 0}(P_{i} - P_{i}',\rp_{j} - \rp_{j})  \ .
\end{eqnarray}
Having now determined the initial condition
Green function,
we focus on the calculation of the medium.
Using the linearity of the free field equations
and the Fourier transform we can subtract
all vacuum contributions from $G_{0}(P^{i},\rp^{j})$
to obtain ${\tilde{G}}_{0}(P^{i},\rp^{j})$.
By making a high-temperature expansion
\begin{equation}
g_{\small BE}(\omega) \approx \frac{1}{\beta \omega} \ , \ (\beta \omega ) << 1
,
\end{equation}
these integrals can be calculated
analytically.
Repeating the steps leading from
Eq.(\ref{eq:genwig}) to Eq.(\ref{eq:genres})
yields the general solution.
In the case under consideration
inhomogeneities only exist in the $X^{3}$
direction, so the dependence on $P_{\perp}$
is simply a $\delta (P_{\perp})$.
The measure of off-shellness now takes the form
\begin{equation}
\alpha = \sqrt{ \rp^{2} - m^{2} } \ ,
\end{equation}
and the boostfactor is
\begin{equation}
\gamma = \frac{1}{ \sqrt{1 - \frac{ \rp_{3}^{2} }{\rp_{0}^{2} } }} \ .
\end{equation}
The solution thus has a simpler form
\begin{equation}
\label{eq:sol}
g(X,\rp ) = \frac{1}{\gamma \alpha \rp^{0}} \{ ( A ) \sin \gamma \alpha (
X^{3} - \frac{\rp^{3}}{\rp^{0}} X^{0} ) + ( B ) \cos \gamma \alpha (
X^{3} - \frac{\rp^{3}}{\rp^{0}} X^{0} ) \} \ ,
\end{equation}
where the A and B are given by
\begin{eqnarray}
A & = & -\frac{\pi T}{\gamma \alpha} \frac{ \rp_{0}^{2} -
\frac{1}{4} \gamma^{2}\alpha^{2} }{\rp_{0}^{4} -
\gamma^{2}\alpha^{2}\rp_{3}^{2}} \nonumber \\
B & = & - \pi T \frac{\sqrt{\rp_{0}^{2} - \rp_{3}^{2} -
\frac{1}{4}\gamma^{2}\alpha^{2} } }
{\rp_{0}^{4} - \gamma^{2}\alpha^{2}\rp_{3}^{2} }  \ .
\end{eqnarray}
If one averages Eq.(\ref{eq:sol})
over small segments in $X^{3}$,
or in $\rp^{3}$,
the $\alpha = 0$ poles become stronger
by a factor $\alpha^{-2}$.
This suggests that the semi-classical
limit may be uncovered not by setting
$\alpha = 0$,
but rather by some kind of averaging.
Since the propagation of the waves
appears to be dominated by the
pole at $\alpha = 0$,
this suggests averaging over the
off-shellness.
In the general case this could replace an
integration over $P_{\perp}$,
but here there is no non-trivial $P_{\perp}$
dependence.
Consequently we will integrate over the
momenta perpendicular to the $\rx^{3}$
axis when averaging over $\alpha$.
The integration over $\alpha$
is restricted to the range
\begin{equation}
0 \leq \alpha \leq \sqrt{ \rp_{0}^{2} - \rp_{3}^{2} - m^{2} }
\end{equation}
due to the reality condition on $\alpha$.
Using the freedom to make a Lorentz transform
to the $\rp^{3}$ frame,
one obtains
\begin{eqnarray}
g(X_{3},\rp_{0}) & = & 4\pi \frac{T}{\rp_{0}}
\{ \frac{ \frac{\pi}{2} - Si ( 2 \sqrt{ \rp_{0}^{2} - m^{2} } X_{3} )}{
\rp_{0}^{2} } \nonumber \\
& + & \frac{1}{4\rp_{0}^{4}X_{3}^{2}} [2 \sqrt{ \rp_{0}^{2} -
m^{2} } X_{3} \cos (2 \sqrt{ \rp_{0}^{2} - m^{2} } X_{3} )
+ \sin (2 \sqrt{ \rp_{0}^{2} - m^{2} } X_{3} ) ] \nonumber \\
& + & \frac{ 2j_{-1}( \frac{ \sqrt{\rp_{0}^{2}-m^{2}}}{\rp_{0}} , \rp_{0} X_{3}
) }{X_{3}\rp_{0}^{2} }\} \ ,
\end{eqnarray}
where we have defined the function $j_{-\nu}(x,y)$ by
\begin{equation}
j_{-\nu}(x,y) = \int_{0}^{x} dz \ (\frac{1}{2} y)^{\nu} ( \sqrt{1 - z^{2}
})^{\nu - \frac{1}{2}}\ \cos(yz) \ .
\end{equation}
with $\nu = 1$.
The function $g(X_{3},\rp_{0})$ is related to
the non-equilibrium generalization $f(X_{3},\rp_{0})$
of the Bose-Einstein distribution $g_{\small BE}$ by
\begin{equation}
f(X_{3},\rp_{0}) = \rp_{0}^{2} g(X_{3},\rp_{0}) \ ,
\end{equation}
except for an overall normalization.
The distribution function $f(X_{3},\rp_{0})$
is represented graphically
in fig.2 as a function of $X_{3}$ for
fixed $\rp^{0}$, and in fig.3 as a
function of $\rp^{0}$ for fixed $X_{3}$.
Two interesting limiting cases
that one may consider are
$\rp_{0} \rightarrow \infty$,
and $X_{3} \rightarrow 0$.
At increasing energy,
$f(X_{3},\rp_{0})$,
behaves more and more like a step
function being zero in $M_{+}$
and thermal in $M_{-}$.
This is the result
that one expects classically.
The second limit, $X_{3} \rightarrow 0$,
represents a purely field theoretical
effect.
The divergence near the
Dirichlet wall is nothing
but the one-plate Casimir effect.
In the presence of such a wall,
quantum fluctuation near it will
be modified.
The modification of vacuum fluctuations
gives rise to a $1/X_{3}^{2}$
behaviour \cite{Full}.
On dimensional grounds one
expects the finite temperature
contribution to be proportional
to $T/X_{3}$, which is the
behaviour we find.
An explicit calculation verifies this
interpretation.
As the wall is instantaneously
removed at $t=0$,
this leads to particle creation.
It is exactly this type of effect
that the SO(1,1) Ward identities
for ICSB adress.

\section{Conclusions}

\indent
In principle a kinetic field theory of
non-equilibrium physics is viable
beyond a two-scale approach.
It is the ideal starting point for
treating inhomogeneous systems
out of equilibrium.
In particular we have studied a
transport theory for the scalar
$\lambda \varphi^{4}$ field theory,
constructed using the closed time
path formalism.
In this case we find that energy
momentum conservation
is guaranteed by the dynamics of
the 2-point functions.
Gradient expansions occurring
naturally within the formalism
cannot be truncated without
either violating energy-momentum
conservation,
or introducing non-locality.
It has been shown how the broken SO(1,1)
symmetry of non-equilibrium field theory
is useful in establishing a physical
interpretation for $G^{ab}$.
Temporal current non-conservation combined
with energy-momentum conservation
determines the ultra-violet behaviour
of the dressed propagators.
The free field equations have been
analyzed on their non-equilibrium
content.
Inhomogeneity and off-shellness
are intimately related in this
context.
This is not expected to change
in the interacting theory.
By integrating out the
off-shell contributions one
arrives at quasi classical
distribution functions.

\bigskip

{\bf Acknowledgements }
\par
We are grateful to J. H\"ufner for his continuing
encouragement and discussions during the course
of this work.
One of us, F.W,
would like to thank
C.\ Villarreal for useful discussions on
the Casimir effect at finite temperature.

\newpage

{\bf Figure Captions}

\begin{itemize}

\item{Fig. 1a and 1b:} The diagrammatical representations of Eq.(2.24) and
Eq.(2.25) are given. The shaded triangle denotes the 1PI 3-vertex
$\alpha_{abc}^{3}$, an open triangle refers to the general 3-point Green
function ${\cal{G}}^{abc}$, thin solid lines represent outgoing lines, single
blobs correspond to the mean fields \mfi{a}, and lines with a blob along the
line represent the full propagator $G^{ab}$. Additionally in 1b the shaded disc
is the 1PI 4-vertex $\alpha_{abcd}^{4}$ and the open disc the general Green
function ${\cal{G}}^{abcd}$.

\item{Figures 2 a-d:} The distribution function $f(x,e)$ as a function of the
energy $e=\rp_{0}$ at $x= -5, -1, 1, 5$ fm/c. Here the temperature $T=2$ GeV
and the mass of the bosons was set $m=0.2$ GeV.

\item{Figures 3 a-d:} The function $f(x,e)$ as a function of $x$ at $e = 0.2,
0.3, 1, 2$ GeV.

\end{itemize}

\end{document}